\documentclass{aa}  

\usepackage{graphicx}
\usepackage{xcolor}
\usepackage[colorlinks=true, citecolor=blue]{hyperref}
\usepackage{txfonts}
\DeclareUnicodeCharacter{2212}{-}

\begin{document}

   \title{Challenges of standard halo models in constraining galaxy properties from CIB anisotropies}

   \author{Athanasia Gkogkou\inst{1,2}
            \and
             Guilaine Lagache\inst{2}
             \and
             Matthieu B\'ethermin\inst{2, 3}
            \and 
            Abhishek Maniyar\inst{4,5}
        }

   \institute{Institutes of Computer Science and Astrophysics, Foundation for Research and Technology Hellas (FORTH), Greece \email{agkogkou@ia.forth.gr}
    \and
     Aix Marseille Univ, CNRS, CNES, LAM, Marseille, France 
     \and
      Université de Strasbourg, CNRS, Observatoire astronomique de Strasbourg, UMR 7550, 67000 Strasbourg, France
      \and
      Kavli Institute for Particle Astrophysics and Cosmology, Stanford University, 452 Lomita Mall, Stanford, CA 94305, USA
      \and
      SLAC National Accelerator Laboratory, 2575 Sand Hill Road, Menlo Park, CA 94025, USA
         }

   \date{Received 07/07/2025; accepted 19/09/2025}


\abstract
{The halo model, combined with halo occupation distribution (HOD) prescriptions, is widely used to interpret cosmic infrared background (CIB) anisotropies and extract physical information about star-forming galaxies and their connection to large-scale structure. Recent CIB-specific implementations of the halo model have adopted more physical parameterizations. However, the extent to which these models can reliably recover meaningful physical parameters remains uncertain. We assess whether the current parameterization of CIB halo models is sufficient to recover astrophysical quantities, such as star formation efficiency, $\eta(M_h, z)$, and halo mass at which the peak of star formation efficiency occurs, $M_{\rm max}$, when fitted to mock data. Also whether discrepancies arise from assumptions about galaxy emission (the HOD ingredients) or from more fundamental components in the halo model, such as bias and matter clustering. We fit the M21 CIB HOD model, implemented within the halo model framework, to mock CIB power spectra and star formation rate density (SFRD) data generated from the SIDES-Uchuu simulation, and compare the best-fit parameters to the known simulation inputs. We then repeat the analysis using a simplified version of the simulation (SSU), explicitly designed to match the HOD assumptions. A detailed comparison of model and simulation outputs is carried out to trace the origin of observed discrepancies. While the M21 HOD model provides a good fit to the mock data, it fails to recover the intrinsic parameters accurately, particularly the halo mass at which star formation efficiency peaks. This mismatch persists even when fitting data generated with the same model assumptions. We find strong agreement (within 5\%) in the emission-related components (SFRD, emissivity), but observe a scale- and redshift-dependent offset exceeding 20\% in the two-halo term of the CIB power spectrum. This likely arises from limitations in the treatment of halo bias and matter clustering within the linear approximation. Additionally, incorporating scatter in the SFR–halo mass relation and the spectral energy distribution (SED) templates significantly affects the shot noise ($\sim 50\%$), but has only a modest impact (less than 10\%) on the clustered component. These results suggest that recovering physical parameters from CIB clustering requires improvements to the cosmological ingredients of the halo model framework, such as adopting scale-dependent halo bias and nonlinear matter power spectra in addition to careful modeling of emission physics.}


\titlerunning{CIB modeling with HOD}
\authorrunning{A. Gkogkou et al.} 

   \maketitle
%

\section{Introduction}

The cosmic infrared background (CIB) is the second most prominent cosmic background. It originates from emissions of dust-enshrouded galaxies undergoing star formation. Studying it helps understanding the cosmic star formation history up to high redshifts, i.e., $z \sim 6$ \citep[e.g.,][]{gispert2000, maniyar2018, jego2023}. Over the years, high-precision measurements of CIB anisotropies have been achieved \citep{lagache2007, planckcollaboration2011, Thacker2013, viero2013, planckcollaboration2014, viero2019}, with additional advancements expected from upcoming missions \citep{zemcov2018, matsuura2024}. These efforts have offered insights into the spatial distribution of the numerous unresolved extragalactic sources. We can thus probe the dusty star-forming galaxy distribution within the large-scale structures of the Universe and connect it to the underlying dark matter.

For the interpretation of the CIB anisotropies, the halo model introduced by \cite{cooray2002} has been extensively employed in the literature \citep[see][for a review]{asgari2023}. This model is based on the assumption that the dark matter is collapsed into symmetric halos and that all galaxies reside within these halos. Its key components include the halo mass function (HMF), halo bias quantifying the over-density or under-density of dark matter halos compared to the overall matter distribution, halo density profile, and halo occupation distribution (HOD). Together, these ingredients describe the abundance, clustering, internal structure of halos, and galaxy properties within them. 

Initial HOD models connected total CIB emissivity solely with the power spectrum of the galaxy number density contrast \citep{viero2009, planckcollaboration2011, amblard2011}, evolving to models considering the luminosity of galaxies dependent on host dark matter halo mass \citep{shang2012, viero2013, planckcollaboration2014,zagatti2024}. When fitting a model to the available CIB power spectrum data, there is a need to retrieve key quantities relevant to galaxy evolution, such as the star formation rate density (SFRD). In this context, \cite{bethermin2013} introduced a connection between galaxy star formation rate (SFR) and baryonic accretion rate (BAR) onto the host dark matter halos. Expanding on this, \cite{maniyar2021} (hereafter M21) maintained this link to propose a four-parameter HOD model. This model successfully fits CIB power spectrum data from \textit{Planck} and \textit{Herschel}, while also aligning with SFRD available data, succeeding in extracting high-z star formation rate (SFR) information from the CIB power spectrum.

While these models adequately fit observational data, uncertainties persist regarding their obtained parameters. For instance, estimates of the halo mass hosting the most efficient star formation range from $10^{12.1} \, M_{\odot}$ as estimated by \cite{viero2013} to $10^{12.94} \, M_{\odot}$ in the latest model by \cite{maniyar2021}. This discrepancy emphasizes the need for validation of the standard HOD models within the halo model framework. Such validation is crucial not only for interpreting CIB data but also for forecasting line intensity mapping (LIM) experiments, which employ similar methodologies but focus on narrower redshift slices \citep{kovetz2017, bernal2022}. To address this, we aim to validate the latest HOD model from \cite{maniyar2021} by comparing it to realistic simulations, advancing towards a new generation of more accurate and precise HOD models for the CIB.

The paper is structured as follows: In Sect.\,\ref{sect:methods}, we describe both the SIDES-Uchuu simulation and the HOD model employed in our analysis. Sect.\,\ref{sect:fit2realSIDES} presents results from fitting the HOD model to mock SIDES-Uchuu data, simulating \textit{Planck} High Frequency Instrument (HFI) data. This analysis concentrates on \textit{Planck} frequencies, offering a detailed exploration of high-z effects, though it is also applicable to \textit{Herschel} data. Sect.\,\ref{sect:hod_vs_simplifiedSIDES} identifies discrepancies within the halo model framework by fitting to mock data from a simplified SIDES-Uchuu (SSU) simulation version, closely aligning with HOD model assumptions and parameterization. Sect.\,\ref{sect:compare_HOD_simu_cosmology-driven} delves into the sources of discrepancy originating from fundamental halo model components. In Sect.\,\ref{sect:scatter_effect} we explore the impact of excluding scatter in astrophysical recipes within the HOD model. Finally, we summarize our results and present the conclusions in Sect.\,\ref{sect:conclusions}.

\section{Method}
\label{sect:methods}

\subsection{SIDES-Uchuu simulation}
\label{subsect:simulations}

The SIDES-Uchuu simulation is the product of two distinct simulations, namely the Simulated Infrared Dusty Extragalactic Sky (SIDES)\footnote{\url{https://cesamsi.lam.fr/instance/sides/home}} and the Uchuu N-body simulation from the Uchuu suite\footnote{\url{http://www.skiesanduniverses.org/Simulations/Uchuu/}}. The combination of these simulations is detailed in \cite{gkogkou2023}. SIDES simulates the far-infrared (FIR) and submillimeter sky by using observed empirical relations \citep{bethermin2017, bethermin2022}. It populates the dark-matter halos initially provided by the Uchuu N-body cosmological simulation with galaxies of specific stellar masses through abundance matching \citep{gkogkou2023}.

The Uchuu cosmological simulation \citep{ishiyama2021}, provides a cosmological box with a comoving volume of 2000 $\rm h^{-1}Mpc$ and a mass resolution of $ \rm 3.27 \times 10^8 \, h^{-1}M_{\odot}$ (with halos resolved at $\gtrsim \, 40$ particles). The resulting lightcone spans nearly 6 orders of magnitude in halo masses at z=0, gradually decreasing to 3-4 orders of magnitude by z=7. The completeness of the Uchuu lightcone down to approximately $\rm \sim 1.3 \times 10^{10} \, h^{-1}M_{\odot}$ across all redshifts is established by comparing halo mass functions (HMFs) with theoretical predictions from \cite{despali2016a}. The Uchuu simulation adopts cosmological parameters consistent with \cite{planckcollaboration2020}.

In the SIDES framework the stellar mass regulates various galaxy properties (e.g., SFR, $L_{\rm IR}$) using a population of main-sequence galaxies \citep[e.g.,][]{daddi2007, schreiber2015} and starburst galaxies \citep{sargent2012, bethermin2012}. There is also a fraction of passive galaxies that are assumed to be non-emitting but still populate the dark matter halos. Additionally, far-infrared and submillimeter line emissions are generated based on this framework \citep{bethermin2022}.

In the simulation, galaxy SFR is determined based on its stellar mass and redshift, with values drawn from the distributions of main sequence and starburst galaxies as described in \cite{schreiber2015}, including a scatter of 0.3\,dex. $L_{\rm IR}$ values are derived from SFR values using the relation established by \cite{kennicutt1998}. Spectral energy distribution (SED) templates are selected from the library provided by \cite{magdis2012} and updated at $z>2$ in \cite{bethermin2015} and \cite{bethermin2017}. In addition, a scatter of 0.2 dex on the mean intensity of the radiation field ($\langle U \rangle$) is implemented in the model, a parameter strongly correlated with dust temperature and influencing the SED template selection for each galaxy. All sources are assigned a magnification value ($\mu$), resulting in either strong or weak lensing.

\subsection{CIB halo model}
\label{subsect:hod_model}

The CIB comes from the dust emission from all the star-forming galaxies across cosmic time; it can thus probe the large-scale structures and provide important information about the clustering of the galaxies. In order to study the large-scale fluctuations of the CIB (or any other large-scale tracer with a specified halo profile), a halo model has been introduced \citep[for a review][]{cooray2002,asgari2023}. In this framework the underlying assumptions are: 1) all the dark matter is enclosed in collapsed dark matter halos, 2) all the galaxies lie in the dark matter halos, and 3) the clustering of the galaxies can be described by the sum of two components: the one- and two-halo terms. The former describes the correlation of the galaxies within a single halo and the latter describes the correlation of the galaxies within different halos. The required ingredients for such a model are: a halo mass function (HMF) that specifies the number density of dark matter halos of a given mass, a halo bias model that quantifies the degree to which the distribution of dark matter halos deviates from that of the underlying matter distribution, a halo profile specifying the distribution of the matter within the halos, and a halo occupation distribution (HOD) prescription that defines how dark matter halos are populated with galaxies. More specifically in the CIB studies, it determines how the light, often quantified as emissivity, is distributed within these dark-matter halos.

\cite{maniyar2021} (hereafter M21) have developed and presented a halo model for the CIB fluctuations. This model links the baryonic matter accretion rate (BAR) onto the dark matter halos to their total star formation rate (SFR). The BAR is defined by the halo mass at a given redshift as
\begin{equation}
\centering
    BAR(M_h,z) = \langle \Dot{M} (M_h, z) \rangle \times \frac{\Omega_b(z)}{\Omega_m(z)},
    \label{eq:bar}
\end{equation}
where $\Omega_b$ and $\Omega_m$ are the dimensionless cosmological baryonic density and total matter density, respectively, and $\Dot{M}(M_h, z)$ is the mean mass growth rate, taken from \cite{fakhouri2010} and is defined as
\begin{equation}
    \langle \Dot{M} \rangle = 46.1 \left( \frac{M_h}{10^{12}M_{\odot}} \right)^{1.1} \times \left( 1 + 1.11z \right) \sqrt{\Omega_m (1+z)^3 + \Omega_{\Lambda}}.
    \label{eq:mass_growth_rate}
\end{equation}
The SFR is connected to the BAR assuming that the stars are formed from the accreted gas with a certain efficiency. This efficiency is parameterized with a lognormal function
\begin{equation}
\begin{split}
    {\rm SFR}(M_h, z) & = \eta(M_h, z) \times {\rm BAR}(M_h, z) \\
    & = \eta_{\rm max} \, e^{- \frac{\left[ ln(M_h) - ln(M_{\rm max}) \right]^2}{2\sigma^2_{M_h}(z)}} \times {\rm BAR}(M_h, z),
    \label{eq:efficiency}
\end{split}
\end{equation}
where $M_{\rm max}$ is the halo mass with the highest star formation efficiency $\eta_{\rm max}$. $\sigma_{M_{\rm h}}(z)$ is the lognormal variance, which characterizes the halo mass range with efficient star formation and evolves with redshift as,
\begin{equation}
    \sigma_{M_{\rm h}} (z) = \sigma_{M_{\rm h0}} - \tau \times {\rm max}(z - z_c).
    \label{eq:sigma_z}
\end{equation}
$z_c$ is the pivot redshift. Above $z_c$ the width equals $\sigma_{M_{h0}}$, below it, $\sigma_{M_{h}}$ varies with redshift, and $\tau$ sets the strength of that evolution. The selection of the lognormal parameterization served the purpose of regulating the star formation efficiency in low and high halo masses. It has been shown by several studies \citep[e.g.,][]{silk2003,keres2005,bethermin2013} that the efficiency peaks at $\sim 10^{12}-10^{13} \, M_{\odot}$ and significantly drops below and above this range. Different parameterizations have been proposed, as for instance from \cite{moster2013}, who proposed a double power-law. This parameterization was tested in Eq.\,\ref{eq:efficiency} but led to an unrealistically low contribution from the halos above the mass of maximum efficiency at high redshift, as explained in the appendix of \cite{maniyar2021}. Therefore, the lognormal parameterization was used since it offered both a good fit to the observational data and physically accepted results.

The M21 model accounts for the quenching of the satellite galaxies by computing the SFR for the subhalos in two different ways. The first one simply follows Eq.\,\ref{eq:efficiency}, where $M_{\rm h}$ is substituted by the mass of the subhalo. And the second way is by weighting the SFR of the host halo by the ratio of the subhalo mass over the host halo mass ($\rm SFR_{sub} = SFR_c \times (m_{sub} / M_h)$). The subhalo's final SFR is determined by selecting the lower value of the two.

Knowing the SFR of both the main halos and their corresponding subhalos at any given redshift, it becomes possible to calculate the differential emissivity ($\frac{dj_{\nu}}{dlogM_h}$). This quantity represents the emissivity per halo mass bin. For the main halos it is
\begin{equation}
    \frac{dj_{\nu,c}}{dlogM_h}(M_h,z) = \frac{d^2N}{dlogM_h dV} \times \chi^2 (1+z)\times \frac{\rm SFR_c}{K}\times S^{\rm eff}_{\nu}(z) \,,
    \label{eq:diff_emissivity_centrals}
\end{equation}
where $\frac{d^2N}{dlogM_h dV}$ is the halo-mass function, taken from \cite{tinker2008}, and $\chi(z)$ is the comoving distance to redshift $z$. $S^{\rm eff}_{\nu}(z)$ is the effective SED of the infrared galaxies at a given redshift for a given frequency \citep{bethermin2013}. $\rm SFR_c$ is the SFR for the central galaxies with a given halo mass. K is the Kennicutt constant (K = SFR/L$_{\rm IR}$), which has a value of $10^{−10} M_{\odot} yr^{-1} L^{-1}_{\odot}$
for a Chabrier IMF, and $L_{\rm IR}$ is the infrared luminosity (8-1000 $\mu$m). The differential emissivity for the subhalos is computed as
\begin{equation}
\begin{split}
    & \frac{dj_{\nu,sub}}{dlogM_h}(M_h,z) = \frac{d^2N}{dlogM_h dV} \times \chi^2(1+z) \times \\
    & \int \frac{dN}{dlogm_{\rm sub}}(m_{\rm sub} | M_h) \frac{\rm SFR_{sub}}{K} \times S^{\rm eff}_{\nu}(z) \times dlogm_{\rm sub} \,,
\end{split}
\label{eq:diff_emissivity_subs}
\end{equation}
where $\frac{dN}{dlogm_{\rm sub}}(m_{\rm sub}|M_h)$ is the subhalo mass function for the satellite galaxies with a subhalo mass $m_{\rm sub}$ in a main halo of mass $M_{\rm h}$. The subhalo mass function is the one from \cite{tinkerwetzel2010}\footnote{All the results published in M21 were obtained using the astroph version of \cite{tinkerwetzel2010}. It should be noted that this version contains a typographical error in the subhalo mass function, where a value of 0.13 was used instead of the correct value, which is 0.3. In this work we have used the correct one.}. The effective SEDs for the satellite galaxies are assumed to be the same as those of the central galaxies.

Finally, the differential emissivity of the main halos and their subhalos are used to construct the one- and two-halo term along with the other halo model ingredients mentioned previously. The one-halo term of the CIB power spectrum describes the clustering of the galaxies within a single halo and is defined as
\begin{equation}
\begin{split}
    C^{1h}_{\ell,\nu,\nu^{\prime}} & = \int \int \frac{d\chi}{dz} \left( \frac{a}{\chi} \right)^2 \left[ \frac{dj_{\nu,c}}{dlogM_h} \frac{dj_{\nu^{\prime},sub}}{dlogM_h} u(k,M_h,z) \right. \\
    & + \frac{dj_{\nu^{\prime},c}}{dlogM_h} \frac{dj_{\nu,sub}}{dlogM_h}  u(k,M_h,z) \\
    & + \left. \frac{dj_{\nu,sub}}{dlogM_h} \frac{dj_{\nu^{\prime},sub}}{dlogM_h} u^2(k,M_h,z) \right] \left( \frac{d^2N}{dlogM_h dV} \right)^{-1} dz dlogM_h\,,
\end{split}
\label{eq:one-halo_term}
\end{equation}
where $a$ is the scale factor of the Universe, $u(k,M_h,z)$ is the Fourier transform of the halo density profile, which describes the density distribution inside the halo. A Navarro-Frenk-White \citep[NFW,][]{navarro1997} profile was considered. The two-halo term describes the correlation between two galaxies in two different halos of mass $M_{\rm h}$ and $M^{\prime}_{\rm h}$ and is calculated as
\begin{equation}
\begin{split}
    C^{2h}_{\ell,\nu,\nu^{\prime}} & = \int \int \int \frac{d\chi}{dz} \left( \frac{a}{\chi} \right)^2 \left[ \frac{dj_{\nu,c}}{dlogM_h} + \frac{dj_{\nu,sub}}{dlogM_h} u(k,M_h,z) \right] \\
    & \times \left[ \frac{dj_{\nu^{\prime},c}}{dlogM^{\prime}_h} + \frac{dj_{\nu^{\prime},sub}}{dlogM^{\prime}_h} u(k,M_h,z) \right] \\
    & \times b(M_h,z) b(M^{\prime}_h,z) P_{\rm lin}(k,z) d {\rm log} M_h d {\rm log} M^{\prime}_h dz\,,
\end{split}
\label{eq:two-halo_term}
\end{equation}
where b($M_{\rm h}$,$z$) is the halo bias prescription given by \cite{tinker2010}, $P_{\rm lin}(k, z)$ is the linear matter power spectrum.

The clustering of the galaxies can be described by the sum of the one-halo ($C_{\ell}^{\rm 1h}$) and two-halo ($C_{\ell}^{\rm 2h}$) term. There is additionally a shot-noise term ($C_{\ell}^{\rm shot}$), which is a flat component and comes from the fact that the number of sources within a certain surveyed area follows the Poissonian statistics. This component is not predicted by the M21 model, it is rather included as a constant term. Therefore, the total CIB power spectrum is computed as
\begin{equation}
\begin{split}
    C^{\rm CIB,tot}_{\ell,\nu,\nu^{\prime}} & = C^{\rm CIB,clust}_{\ell,\nu,\nu^{\prime}} + C^{\rm CIB,shot}_{\ell,\nu,\nu^{\prime}} \,, \\
    & = C^{\rm 1h}_{\ell,\nu,\nu^{\prime}} + C^{\rm 2h}_{\ell,\nu,\nu^{\prime}} + C^{\rm CIB,shot}_{\ell,\nu,\nu^{\prime}} \,,
\end{split}
\label{eq:total_Cl_CIB}
\end{equation}
where the one-halo term arises from galaxy pairs within the same halo, the two-halo term from pairs in different halos, and the shot-noise term is flat in $\ell$ and captures the Poisson contribution from unresolved sources.

The novelty of the M21 model can be summarized in the following aspects. It is a simple model with only four free parameters ($M_{\rm max}$, $\eta_{\rm max}$, $\sigma_{M_{h0}}$, $\tau$) as opposed to other models with a larger number of parameters \citep[e.g.,][]{viero2013}. By connecting the matter accretion onto the dark matter halos to their SFR, it provides a more physically-driven parameterization compared to other models that assume a luminosity-halo mass relation \citep[e.g.,][]{shang2012,debernardis2012,planckcollaboration2014a,mccarthy2021}. The model was fitted to CIB angular power spectrum data measured by \textit{Planck}, while also taking into account external observational constraints, such as the SFRD and the mean level of the CIB. With only four free parameters, it both obtained a very good fit and it is compatible with the SFRD and mean CIB data.

\subsection{Philosophy of our work}
\label{subsect:philosophy_of_this_paper}

Halo models are widely used to interpret CIB observational data and infer astrophysical and cosmological parameters. However, the extent to which we can trust the parameters derived from these models remains uncertain. Our work has two main goals. The first one is to examine the effectiveness of the parameterization of the M21 HOD model, the latest halo model used for the CIB analysis, to properly extract astrophysical information. The second objective involves a comprehensive evaluation of the family of halo models as a whole for CIB anisotropies. This entails an investigation into their individual components and their ability to establish a realistic connection between the large-scale structures and the emission from individual DSFGs. Our approach involves comparing the halo model to a simulation, where all components and parameters are known beforehand. For this purpose we use the SIDES-Uchuu simulation, which is modified as needed throughout this paper to facilitate various validation tests.

It is worth stressing that our goal is not to capture the full physical complexity of DSFGs in the real Universe. SIDES-Uchuu itself is based on heuristic prescriptions linking halos to galaxies, and thus inevitably omits some processes. Instead, our objective is to test whether a simplified halo-model parameterization can recover the astrophysical quantities present within this simulation, providing a controlled benchmark where the true inputs are known.

The analysis begins by fitting the halo model to mock data, generated using the original SIDES-Uchuu simulation. Subsequently, we compare the SFR-BAR efficiency curve determined by the best-fit parameters to the one inherent in the simulation used to generate the mock data. This step aims to test the accuracy of the chosen parameterization in the M21 HOD model.

As a follow up test, we fit the halo model to a new set of mock data, generated using a modified version of the SIDES-Uchuu simulation. These modifications align the simulation's prescriptions to those employed in the halo model and they involve two key adjustments. First, the SFR drawing method is adjusted from the redshift-evolved main sequence to the analytic formula given in Eq.\,\ref{eq:efficiency}. Second, instead of using SED templates dependent on both redshift and the galaxy's mean radiation field ($\langle U \rangle$), effective SED templates that depend solely on redshift are incorporated. These modified simulations are referred to as simplified SIDES-Uchuu (SSU). This step aims to assess whether the resulting efficiency now closely matches that of the simulation. However, as we find, persistent offsets between the simulation and the efficiency curves from the M21 HOD model suggested a potential inconsistency not just in the specific HOD model (Eq\,\ref{eq:diff_emissivity_centrals} and Eq.\,\ref{eq:diff_emissivity_subs}) but also in the halo model as a framework (Eq.\,\ref{eq:one-halo_term} and Eq.\,\ref{eq:two-halo_term}).

To delve into the investigation of the halo model, we once again use the SSU. We conduct a thorough comparison of the various quantities used to model the CIB power spectrum (Eq.\,\ref{eq:two-halo_term}) in the halo model with the simulation. This allows to systematically isolate and explore different components of the halo model (e.g., SFR, differential emissivity) as potential sources of discrepancies. Finally, we investigate and quantify the impact of introducing some scatter in various recipes (e.g., SFR) on the obtained power spectrum, shedding light on the importance of modeling the scatter in the employed recipes.

\section{Fitting the M21 HOD model to SIDES-Uchuu mock data}
\label{sect:fit2realSIDES}

In this section, we fit the M21 HOD model\footnote{\url{https://github.com/abhimaniyar/halomodel_cib_tsz_cibxtsz}} to mock data obtained with SIDES-Uchuu. We then present the results and discuss their implications.

\subsection{SIDES-Uchuu mock data}
\label{subsect:realSides_data_preparation}

In order to generate mock CIB data as they would be observed from \textit{Planck}, we used the SIDES-Uchuu simulation. We first generated four maps, corresponding to a simulation area of $117 \, \rm deg^2$, for the four highest \textit{Planck} frequencies (217, 353, 545, and 857\,GHz from the HFI instrument). To fulfill this task, we had to multiply the SEDs by the corresponding \textit{Planck} bandpasses \citep[see Sect. 2.6 from][for more details on the cube/map making]{bethermin2022}. Next, we computed the auto- and cross-power spectra of all the maps using the \texttt{powspec}\footnote{\url{https://zenodo.org/record/4507624\#.YTiIfyZR3mE}} python package. From the resulting auto- and cross-power spectra we kept the data points corresponding to the same $\ell$ multipoles as those in the \textit{Planck} data sets. Data points from SIDES-Uchuu and the M21 HOD model are not color corrected \citep[][Eq.\,A.2]{lagache2020}, since they are already in the same convention. Therefore, we did not apply any color corrections. 

We estimated the covariance matrix (see Appendix\,\ref{ap:covariance_matrix}) using the flat-sky implementation of \texttt{pymaster}\footnote{\url{https://namaster.readthedocs.io/en/latest/index.html}} \citep{alonso2019}. The computation uses four frequency channels (217, 353, 545, and 857 GHz) and eight multipole bins for each power spectrum. As a result, the full Gaussian covariance matrix has dimensions ($8 \times 4$, $8 \times 4$) = (32, 32). The error bars on the simulated data points are derived from the square roots of the diagonal elements of this analytically computed covariance matrix.

To determine the shot noise in each simulated map, we shuffled the sources within the simulated catalog before generating the map. In this way any clustering information is removed. We then computed the power spectrum of the shuffled map, which provided a direct measure of the shot-noise level for all the auto and cross power spectra. Having this information allows us to calculate the correlation matrix ($\mathcal{R_{\rm SN}}$), which is used during the fitting process (see Sect.\,\ref{subsect:fitting_results_realSIDES}). This matrix describes the degree of correlation between the shot noise power of the various frequencies. The analytic formula for calculating the correlation matrix is:
\begin{equation}
    \mathcal{R_{\rm SN}} = \frac{C^{\nu,\nu^{\prime}}_{\ell}}{\sqrt{C^{\nu,\nu}_{\ell} \times C^{\nu^{\prime},\nu^{\prime}}_{\ell}}}.
\label{eq:corr_matrix}
\end{equation} 

The M21 model uses external constrains, such as the mean CIB intensity and the SFRD. We thus had to generate mock data for these two quantities. For the mean CIB intensity (CIB monopole), we generated three maps after applying the three \textit{Herschel}/SPIRE filters accordingly (250, 350, and 500 $\mu$m). We use \textit{Herschel}/SPIRE data for the CIB monopole even though we use \textit{Planck} data for the power spectrum to maintain consistency with the methodology followed by \cite{maniyar2021}. We then computed the mean of the generated maps and converted the units from $\rm Jy \, sr^{-1}$ to $\rm nW \, m^{-2} sr^{-1}$. 
In order to obtain the error bars for each CIB mean, we used the error bars of the \textit{Herschel}/SPIRE data after normalizing accordingly for SIDES, as:
\begin{equation}
    \Delta I^{\rm SIDES}_{\nu} = \frac{\Delta I^{\rm SPIRE}_{\nu}}{I^{\rm SPIRE}_{\nu}} \times I^{\rm SIDES}_{\nu}.
\label{eq:mock_errorbar_computation}
\end{equation}

Finally, we obtained the mock SFRD data by computing the total SFR from the SIDES catalogs and summing the SFR values from all the sources in the given redshift bin. This value was divided by the corresponding comoving volume. From the resulting SFRD curve, we only kept the points at the exact same redshifts as the observational data \citep{madau2014,khusanova2021} and used the same relative error bars.

\subsection{Fitting results}
\label{subsect:fitting_results_realSIDES}

We performed a Markov chain Monte Carlo (MCMC) analysis of the global M21 CIB model parameter space using the Python package ‘emcee’ \citep{foreman-mackey2013}. There are 8 free parameters:
\begin{itemize}
    \item physical model parameters: $M_{\rm h,max}$, $\eta_{\rm max}$, $\sigma_{M_{h0}}$, $\tau$
    \item shot noise: $\rm SN_{217}, SN_{353}, SN_{545}, SN_{857}$
\end{itemize}
The global $\chi^2$ has a contribution from the CIB auto/cross power spectra and the priors imposed by the $\langle I_{\rm CIB} \rangle$ and SFRD external observational constraints. 

In contrast to M21, where measurement uncertainties were assumed to be Gaussian and uncorrelated, we use the full covariance matrix (see Appendix \ref{ap:covariance_matrix}) when computing the $\chi^2$ statistic. This allows us to account for correlations between frequencies and provides a more accurate assessment of the model fit. To establish prior distributions for the eight model parameters, a uniform distribution is adopted, with a predefined range. A comprehensive summary of these priors can be found in Table\,\ref{tab:realSIDES_mcmc_results_and_priors}. In the M21 model, concerning the shot noise, only the auto-power shot noise parameters are considered as free parameters. Meanwhile, for the cross-power shot noise, the model computes the values following the correlation matrix from \cite{bethermin2013_cib_z}, which accounts for the fact that different frequencies can have contributions from overlapping redshifts and thus the same sources. As a result, the shot noise for the cross frequencies is defined as follows:
\begin{equation}
    \rm SN_{\nu,\nu^{\prime}} = \mathcal{R} \times \sqrt{SN_{\nu \nu} \, SN_{\nu^{\prime} \nu^{\prime}}},
    \label{eq:cross_sn_compuation}
\end{equation}
where $\mathcal{R}$ is the correlation matrix we have computed from the simulation (Eq.\,\ref{eq:corr_matrix}).

The priors that have been used for the fit are shown in detail in Table\,\ref{tab:realSIDES_mcmc_results_and_priors} together with the best-fit values. For the four physical parameters the priors are exactly the same as those used in \cite{maniyar2021}. But for the priors of the shot-noise levels we chose a $\pm$20\% range centered on the real value of the shot noise for each frequency. We obtained a global $\chi^2$ of 384 for 80 Planck $C_{\ell}$ points. The $\chi^2$ value obtained in \cite{maniyar2021} for the same amount of points was 113. The difference in the quality of the fit is somewhat surprising but it could be the result of the fact that the M21 analysis used the wrong version of the \cite{tinkerwetzel2010} subhalo mass function, while in our analysis we used the updated version with the corrected value. It could additionally be the result of using different priors for the shot noise parameters. M21 placed broad flat priors on the auto-power shot noise with width of [0.1 - 2] times the value estimated by the \cite{bethermin2012} model. In our analysis we used a width of [0.8 - 1.2] times the value we obtained from the shuffled SIDES-Uchuu maps. Narrowing the range further (e.g., $[0.9,1.1]$) pulls the inferred shot-noise levels closer to the true values but does not improve the recovery of the four astrophysical parameters ($\log M_{\rm h,max}$, $\eta_{\rm max}$, $\sigma_{M_{\rm h},0}$, $\tau$). Conversely, widening the shot-noise priors (e.g., $[0.5,1.5]$) broadens the astrophysical posteriors but does not improve the fit to the large-scale, two-halo–dominated part of the spectra, nor does it reduce the overall $\chi^2$. We therefore retain the informative yet conservative $[0.8,1.2]$ baseline.

\begin{figure*}[h!]
    \centering
    \includegraphics[width=0.9\linewidth]{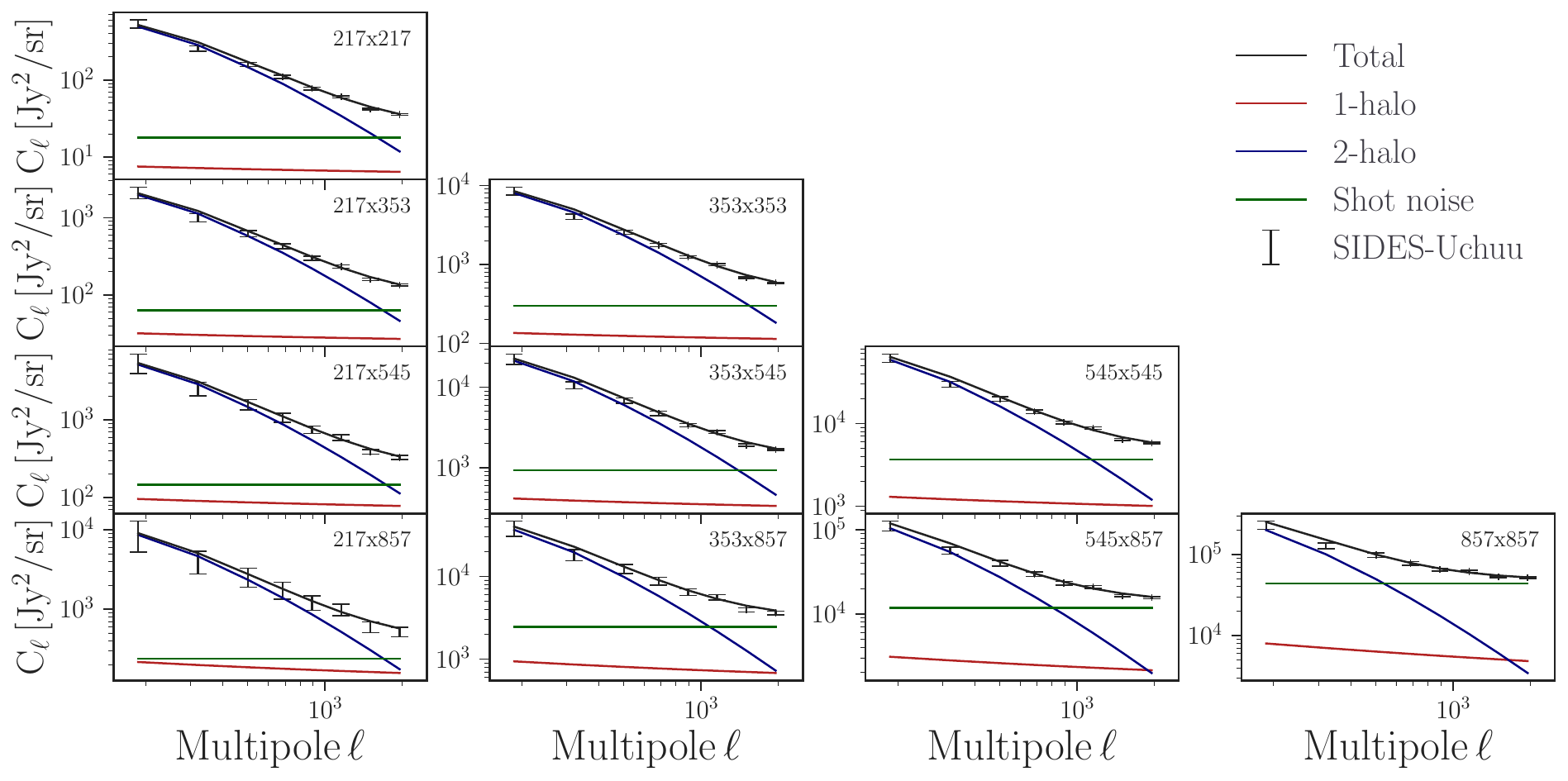}
    \caption{Mock data of the CIB auto- and cross-power spectra obtained by SIDES-Uchuu (black points) and the best-fit CIB halo model (black line) with its components (one-halo term, two-halo term, shot noise). The ratio of the model over the data is shown in Fig.\,\ref{fig:fit2realSIDES_power_spectra_ratios}.}
    \label{fig:fit2realSIDES_Pk}
\end{figure*}

\begin{figure}[h!]
    \centering
    \includegraphics[width=\linewidth]{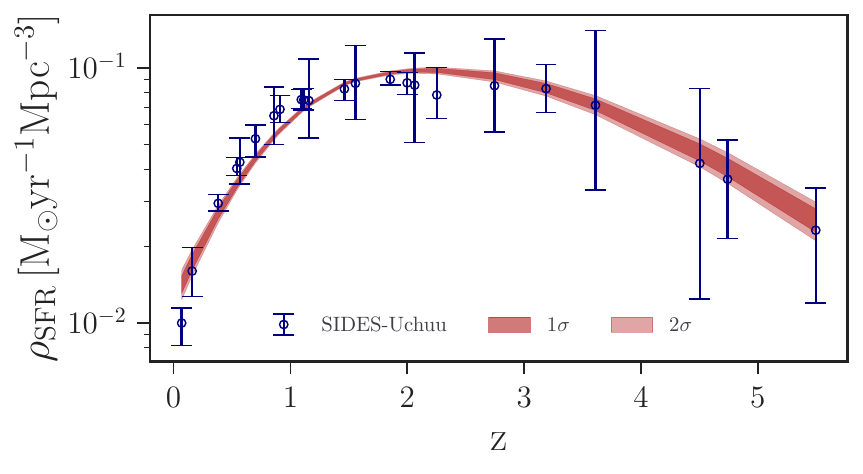}
    \caption{Mock data of the SFRD obtained by SIDES-Uchuu (blue points) and the 1 and 2 $\sigma$ contours of the SFRD models defined by the various sets of parameters explored by the MCMC walkers (red shaded areas).}
    \label{fig:fit2realSIDES_SFRD}
\end{figure}

The resulting best-fit model (one-halo term + two-halo term + shot noise) together with the mock data points are shown in Fig.\,\ref{fig:fit2realSIDES_Pk}. In Fig.\,\ref{fig:fit2realSIDES_SFRD} we show the SFRD data from our simulation and the contours of the multiple models defined by the various sets of parameters explored by the MCMC process. The corner plots of the marginal distributions of the model parameters are displayed in the Appendix\,\ref{ap:complementary_plots} in Fig.\,\ref{fig:fit2realSIDES_cornerplot}.

\begin{table*}[h!]
    \centering
    \begin{tabular}{|c|c|c|c|c|c|c|c|c|}
        \hline
         & ${\rm log}M_{\rm h,max}$ & $\eta_{\rm max}$ & $\sigma_{M_{\rm h0}}$ & $\tau$ & $\rm SN_{217}$ & $\rm SN_{353}$ & $\rm SN_{545}$ & $\rm SN_{857}$ \\
         & $[M_{\odot}]$ & & & & $[\rm Jy^2/sr]$ & $[\rm Jy^2/sr]$ & $[\rm Jy^2/sr]$ & $[\rm Jy^2/sr]$ \\
         \hline
        best-fit value & $12.34^{+0.07}_{-0.08}$ & $0.27^{+0.01}_{-0.01}$ & $1.99^{+0.06}_{-0.12}$ & $1.06^{+0.16}_{-0.18}$ & $17.81^{+0.17}_{-0.32}$ & $300^{+10}_{-11}$ & $3684^{+122}_{-124}$ & $43770^{+887}_{-927}$ \\
        prior lower limit & 8 & 0 & 0 & 0 & 14 & 240 & 2947 & 35016 \\
        prior upper limit & 15 & 1 & 2 & 3 & 21 & 360 & 4421 & 52524 \\
        \hline
    \end{tabular}
    \caption{Best-fit values with uncertainties and the ranges of uniform distributions used as priors for fitting the M21 HOD model to mock data from SIDES-Uchuu.}
    \label{tab:realSIDES_mcmc_results_and_priors}
\end{table*}

\subsection{Discussion}

As evident in Fig.\,\ref{fig:fit2realSIDES_Pk} and supported by the resulting $\chi^2$ value, the M21 HOD model provides a statistically good fit to the mock data. Nonetheless, the goal of this analysis is to draw conclusions regarding the resulting best-fit parameters and their effectiveness in describing the physical quantities embedded in the simulation.

The exact level of the shot noise can be easily computed from the simulated maps we used to obtain the mock data. We can thus directly compare the intrinsic to the best-fit value on a one-to-one basis. Table\,\ref{tab:realSIDES_SN_results} shows both the intrinsic and best-fit values of the four \textit{Planck} frequencies shot noise, along with their corresponding statistical uncertainties. The fractional difference is below 18\%, which is seemingly low, but the best-fit value does not fall within the range defined by the 1$\sigma$-statistical uncertainty for any of the shot noise parameters. This shows the weakness of the fit to recover the shot noise values, potentially due to the strong degeneracy of the shot noise with the one-halo term, as well as the correlation between the various frequencies.

Regarding the ability of the model to reproduce the right SFRD, we can conclude from Fig.\,\ref{fig:fit2realSIDES_SFRD} that the overall shape of the models agrees with that of the mock SFRD. Between $z=0.3$ and $z=1$ there seems to be a systematic underestimate of the models but they still lay within the error bars. At $z > 3$ the range of the potential models increases as well as the size of the error bars. However, the agreement still holds.

\begin{table*}[h!]
    \centering
    \begin{tabular}{|c|c|c|c|c|}
        \hline
         & $\rm SN_{217} \, [\rm Jy^2/sr]$ & $\rm SN_{353} \, [\rm Jy^2/sr]$ & $\rm SN_{545} \, [\rm Jy^2/sr]$ & $\rm SN_{857} \, [\rm Jy^2/sr]$ \\
         \hline
        intrinsic value & 14.88 & 268 & 3745 & 45218 \\
        best-fit value & $17.81^{+0.17}_{-0.32}$ & $300^{+10}_{-11}$ & $3684^{+122}_{-124}$ & $43770^{+887}_{-927}$ \\
        diff (\%) & 19.7 & 11.9 & 1.6 & 3.2 \\
        \hline
    \end{tabular}
    \caption{Intrinsic and best-fit values of the shot noise for the four Planck frequencies (217, 353, 545, 857 GHz).}
    \label{tab:realSIDES_SN_results}
\end{table*}

\begin{figure}[h!]
    \centering
    \includegraphics[width=\linewidth]{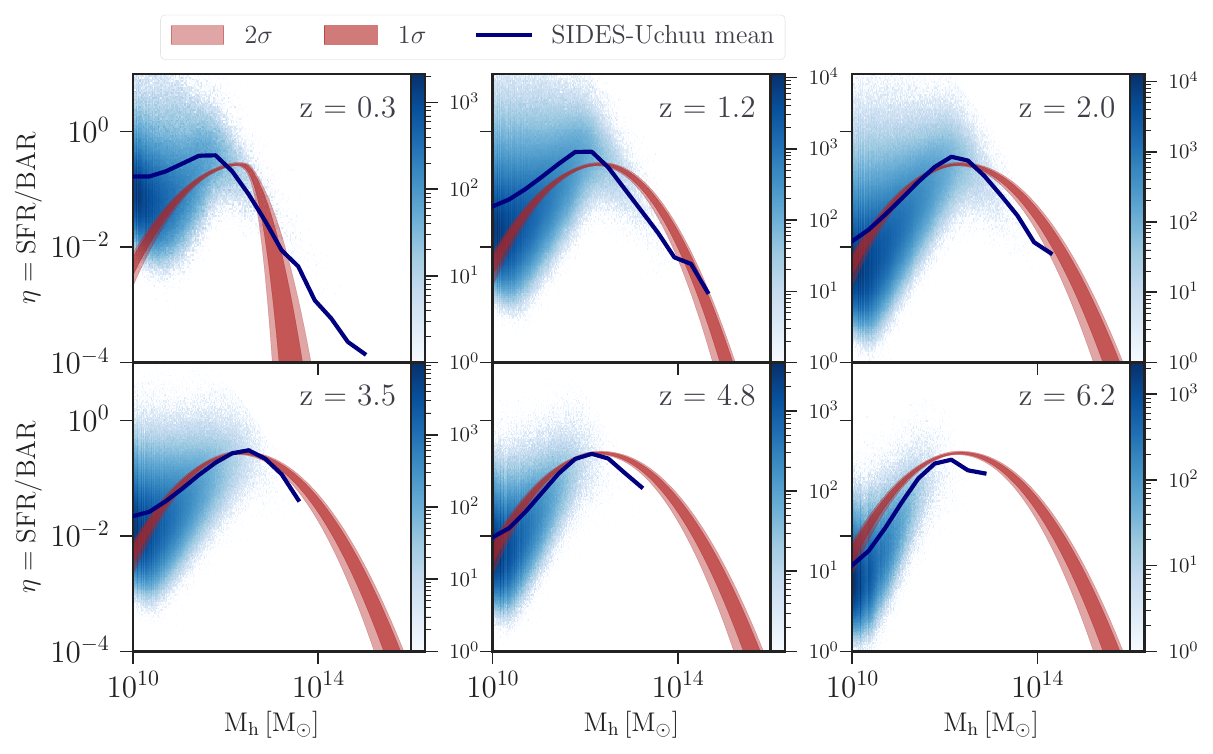}
    \caption{SFR-BAR efficiency from SIDES-Uchuu (blue gradient) and the mean efficiency (blue solid line). The various models from the MCMC parameter exploration of the M21 HOD model is shown in red (red contours for 1$\sigma$ and 2$\sigma$). Each subplot corresponds to a different redshift, with the center of the redshift bin displayed on the top right of each panel. The redshift interval used to compute the density of the SFR/BAR values of the sources in the simulation is $\Delta$z=0.05. The colorbar indicates the SFR/BAR value density, with blue representing high density and white indicating lower density, implying fewer sources with these values.}
    \label{fig:eta_Mh_realSIDES}
\end{figure}

In order to assess the reliability of the remaining four best-fit parameters ($M_{\rm max}$, $\eta_{\rm max}$, $\sigma_{M_{h0}}$, $\tau$), which define the SFR-BAR efficiency, it is essential to compare the simulation efficiency with the efficiency curve defined by the best-fit parameters. This comparison is depicted in Fig.\,\ref{fig:eta_Mh_realSIDES}, where the 1\,$\sigma$ and 2\,$\sigma$ contours of the models are plotted together with the SIDES data points and their mean. The model manages to capture the general shape of the intrinsic SFR efficiency, but falls short in accurately reproducing the amplitude ($\eta_{\rm max}$) and the halo mass of the highest star formation efficiency ($M_{\rm max}$). $M_{\rm max}$ seems to be always overestimated by the halo model. This inconsistency becomes more pronounced at $z < 2$ and can significantly impact the interpretation of the CIB, as the primary flux contribution is expected to originate from $1<z<2$ \citep[e.g., Fig.\,4 from][]{maniyar2018}.

This result could imply that there may be more suitable parameterizations for the efficiency between SFR and BAR that can capture the redshift evolution more accurately while still maintaining the simplicity of having only four free parameters. 

\section{Investigation of discrepancies within the halo model}
\label{sect:hod_vs_simplifiedSIDES}

In this section we aim to investigate in detail the source of the inconsistency we noticed in Sect.\,\ref{sect:fit2realSIDES}. For this purpose, we modify the SIDES-Uchuu simulation by incorporating the same parameterizations as in the M21 HOD model. This allows for a direct comparison.

\subsection{Simplified SIDES-Uchuu (SSU)}
\label{subsect:simplified_SIDES}

In the previous test the halo model failed to reproduce the quantities intrinsic to SIDES-Uchuu. However, there are recipes that are different from those in the halo model. Therefore, we produce an alternative version of SIDES-Uchuu using the same exact parameterization as in the M21 HOD model. In this way we aim to explore potential biases that are not caused by the parameterization of the model. We implemented the following modifications: the SFR recipe, the implemented SED templates for the galaxies, and the incorporation of the quenching process.

{\tt SFR recipe:}
Within the existing version of SIDES, the SFR values are determined based on the redshift-evolving main sequence. To align SIDES with the M21 HOD model, we incorporated in SIDES the SFR parameterization of the HOD model, described by Eq.\,\ref{eq:efficiency}. Consequently, in the simplified version of our simulation, the SFR value is solely determined by the halo mass ($M_{\rm h}$) and the redshift, precisely matching the assumption made in the M21 HOD model. One can choose to simplify this recipe further by disabling the redshift-induced evolution of the efficiency curve's width. This can be achieved by setting $\tau=0$ in Eq.\ref{eq:sigma_z}. This is further investigated in Appendix\,\ref{ap:tau_off_results}. Finally, no scatter is added to the assigned SFR values.

{\tt SED templates:}
To estimate the continuum emission ($L_{\rm IR}$) SIDES adopts a SED template sourced from the updated template library introduced in \cite{bethermin2017}. Each template accounts for the galaxy type (main sequence or starburst), redshift, and the mean radiation field ($\langle U \rangle$) closely linked to the dust temperature. However, in the M21 HOD model in order to determine the emissivity of galaxies within a dark matter halo, an effective SED is assumed. The effective SED at a specific redshift is calculated as the average of all SEDs. These SEDs are from the same library as those used in SIDES \citep{bethermin2017}, corresponding to different $\langle U \rangle$ values for each galaxy type at that redshift. Each SED is weighted based on its contribution to the infrared luminosity density. After obtaining the final effective SED (Eq.\,\ref{eq:diff_emissivity_centrals}), a specific bandpass (e.g., Planck, \textit{Herschel}/SPIRE) is applied. Subsequently, we adopted the same effective SEDs for the simplified version of SIDES.

{\tt Quenching and subhalos:} 
In the M21 HOD model there is a way to account for the quenching as explained in Sect.\,\ref{subsect:hod_model}. This allows for the incorporation of the influence of the subhalo's environment, which can potentially result in a reduced SFR of the subhalos. However, in SIDES all dark-matter halos, whether they are main or subhalos, are treated uniformly. This implies that the same scaling relations are applied without distinguishing between centrals and subhalos. In order to directly compare the model to the simulation, it is more convenient to deactivate the quenching module in the model rather than introducing a new recipe in SIDES specifically designed to account for quenching. It is also possible and straightforward to discard the subhalos contribution, either in the model, the simulation, or both if needed.

{\tt Lensing:}
Finally, in the simplified version of SIDES we switch the lensing module off (i.e. $\mu = 1$) for all the galaxies. This aims to prevent any additional contributions to the tension between the model and the simulation.

\subsection{Fitting process and results}
\label{subsect:fitting_results_simplifiedSIDES}

We generated two mock data sets by using the SSU. We choose the parameters ${\rm log}M_{\rm h,max}$, $\eta_{\rm max}$, $\sigma_{M_{\rm h0}}$, and $\tau$ to be the same as the values obtained in M21 \footnote{The selection of the input parameters is not the primary focus of this analysis. Instead, the goal is to assess whether the model can retrieve the selected input parameters across various scenarios through the fitting process.}. The first data set includes both central dark-matter halos and subhalos (referred to as the "with subs" data set), and the second one is without subhalos (referred to as the "no subs" data set)\footnote{In the "no subs" case, subhalos are excluded from the SSU simulation and in the halo model all ‘sub’ terms are set to zero.}. We have followed the same exercise by setting the $\tau$ parameter to 0 to investigate the effect of an even more simplified parameterization. The results can be found in Appendix\,\ref{ap:tau_off_results}.

Each mock data set comprises of: 1) CIB auto- and cross-power spectrum data points for the four highest Planck frequencies, 2) mean CIB intensity data at the three \textit{Herschel}/SPIRE bandpasses, and 3) SFRD data points within the redshift range of 0 to 5. The error bars for the power spectra are derived from the square root of the diagonal elements of the covariance matrix, as detailed in Appendix\,\ref{ap:covariance_matrix}. The SFRD and mean CIB intensity data points are computed following the same procedure described in Sect.\,\ref{subsect:realSides_data_preparation}.

\begin{figure}[h!]
    \centering
    \includegraphics[width=\linewidth]{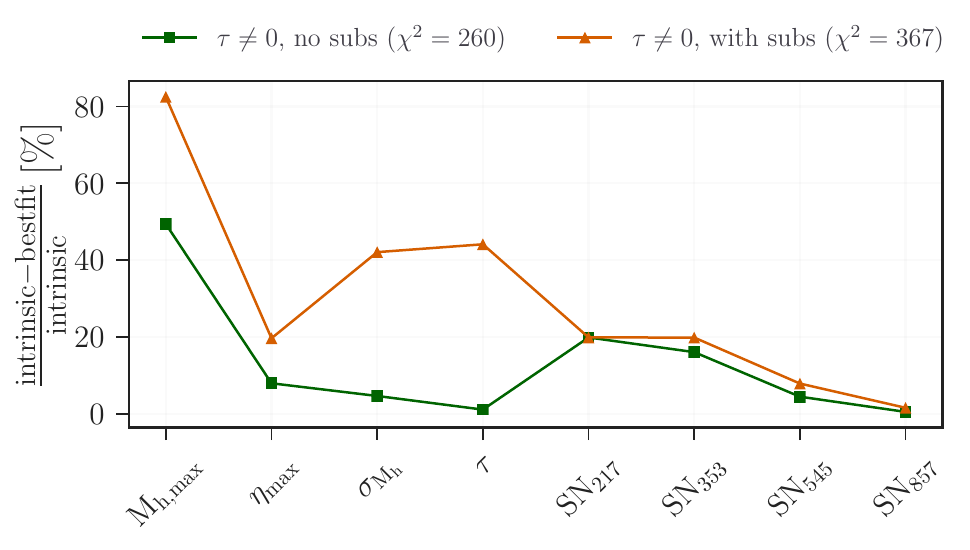}
    \caption{Percentage difference between the intrinsic and the best-fit values of the model parameters from the halo model fit to the SSU simulated data. The green color corresponds to the "no subs" case, and the orange color corresponds to "with subs".}
    \label{fig:perecentage_difference_tauON}
\end{figure}

Similarly to Sect.\,\ref{subsect:fitting_results_realSIDES} we conducted a MCMC analysis for both mock data sets. The obtained best-fit parameters and the global $\chi^2$ values are gathered in Table\,\ref{tab:simplifiedSIDES_mcmc_results}. Comparing the input values to the best-fit results reveals that, although the model provides a good fit to the simulated data, the recovered parameter values deviate significantly from the true inputs. The resulting corner plots are shown in Fig.\,\ref{fig:corner_plot_simplified_with_vs_no_subs_tauON}. We furthermore show in Fig.\,\ref{fig:perecentage_difference_tauON} the fractional difference between the intrinsic and best-fit values for each parameter across the two different data sets.

\begin{table*}[h!]
    \centering
    \begin{tabular}{|c|c|c|c|c|c|}
        \hline
         & \multicolumn{2}{c|}{$\tau = 0$} & \multicolumn{2}{c|}{$\tau \neq 0$ } & intrinsic \\
         & no subs & with subs & no subs & with subs & values \\
         \hline
        ${\rm log}M_{\rm h,max}$ & $12.55^{+0.06}_{-0.06}$ &  $12.08^{+0.06}_{-0.06}$ & $12.64^{+0.05}_{-0.07}$ & $12.18^{+0.04}_{-0.09}$ & 12.94\\
        $\eta_{\rm max}$ & $0.44^{+0.10}_{-0.06}$ &  $0.78^{+0.17}_{-0.17}$ & $0.39^{+0.04}_{-0.03}$ &  $0.50^{+0.14}_{-0.10}$ & 0.42 \\
        $\sigma_{\rm M_{h0}}$ & $1.28^{+0.23}_{-0.25}$ &  $0.53^{+0.22}_{-0.13}$ & $1.67^{+0.18}_{-0.21}$ &  $1.01^{+0.27}_{-0.23}$ & 1.75 \\
        $\tau$ & - & - & $1.16^{+0.14}_{-0.18}$ &  $0.65^{+0.19}_{-0.17}$ & 1.17 \\
        $\chi^2$ & 542 & 290 & 288 & 348 & \\
        \hline
    \end{tabular}
    \caption{Best-fit values together with the uncertainties obtained after fitting the HOD model to the 4 different cases of mock data ($\tau=0$ with and without subhalos, $\tau \neq 0$ with and without subhalos). The intrinsic values of the simulation are indicated in the last column. See Fig.\,\ref{fig:corner_plot_simplified_with_vs_no_subs_tauON} and Fig.\,\ref{fig:corner_plot_simplified_with_vs_no_subs_tauOFF}.} 
    \label{tab:simplifiedSIDES_mcmc_results}
\end{table*}

\subsection{Discussion}
\label{subsect:discussion_fit2simplifiedSIDES}

The M21 HOD model integrated in the halo model framework can fit the mock data, but, similarly to Sect.\,\ref{sect:fit2realSIDES}, the recovered parameters do not exactly match the intrinsic ones, even though the exact same parameterization was used in the SSU (see Fig.\,\ref{fig:perecentage_difference_tauON}). 

In the "with subs" case, the shot noise parameters are recovered within 20\%, but the four astrophysical parameters (${\rm log}M_{\rm h,max}$, $\eta_{\rm max}$, $\sigma_{M_{\rm h0}}$, $\tau$) show discrepancies exceeding 50\%, indicating a poor recovery of the physical quantities that govern the star formation efficiency. This suggests that degeneracies within the model, particularly between the one-halo term and shot noise, may obscure the true values of these parameters. To mitigate this, we fit the halo model to the "no subs" mock data, which excludes subhalos and thus suppresses the one-halo contribution. The fit quality improves, with all parameters recovered within 20\% of their intrinsic values, except for ${\rm log}M_{\rm h,max}$. This supports the hypothesis that reducing the one-halo–shot noise degeneracy improves parameter constraints. However, the persistent offset in ${\rm log}M_{\rm h,max}$, even in the absence of subhalos, suggests that its poor recovery is likely rooted in a more fundamental limitation of the model’s parameterization or in residual degeneracies with other parameters such as $\eta_{\rm max}$ or $\sigma_{M_{\rm h0}}$.

We repeated the above fitting exercise with an even simpler parameterization, setting the $\tau$ parameter to zero.  While some parameters were better constrained, others showed worse results. Overall, this simplification did not lead to significantly better outcomes. The resulting plots are presented and discussed further in Appendix\,\ref{ap:tau_off_results}.

\section{LSS and cosmological components: Impacts on the halo model-simulation tension}
\label{sect:compare_HOD_simu_cosmology-driven}

In the previous sections, we fitted the halo model to mock data generated using both the original and simplified versions of SIDES-Uchuu. However, the obtained fitted values consistently deviated from the intrinsic ones, even when the same parameterization was applied to both the halo model and the simulation. To better understand the origin of these deviations, we now focus on directly comparing the halo model with the SSU simulation for the exact same parameter values.

At various steps of the computation we compare the quantities used by the model to construct the power spectrum. First, the SFRD, next the differential emissivity $\frac{dj}{dlogM_h}$ (Eq.\,\ref{eq:diff_emissivity_centrals}), and finally the power spectra. For all these comparisons, in addition to using the SSU (which shares the same parameterization as the M21 HOD model) we incorporate the exact halo mass function (HMF) from the SIDES-Uchuu simulation into the M21 HOD model. This direct comparison could eliminate potential uncertainties arising from the fitting process (e.g., parameter degeneracies) or differences between the analytical and the Uchuu HMF.

To simplify the analysis and focus on intermediate and large scales, we discard subhalos. This choice is supported by 'no subs' best-fit values that closer align with intrinsic parameters. The intrinsic parameters used in the SSU framework throughout this section correspond to those labeled as `intrinsic' in Table\,\ref{tab:simplifiedSIDES_mcmc_results}.

\subsection{SSU vs HOD model: Star formation rate density (SFRD)}
\label{subsect:sfrd_investigation}

\begin{figure}[h!]
    \centering
    \includegraphics[width=\linewidth]{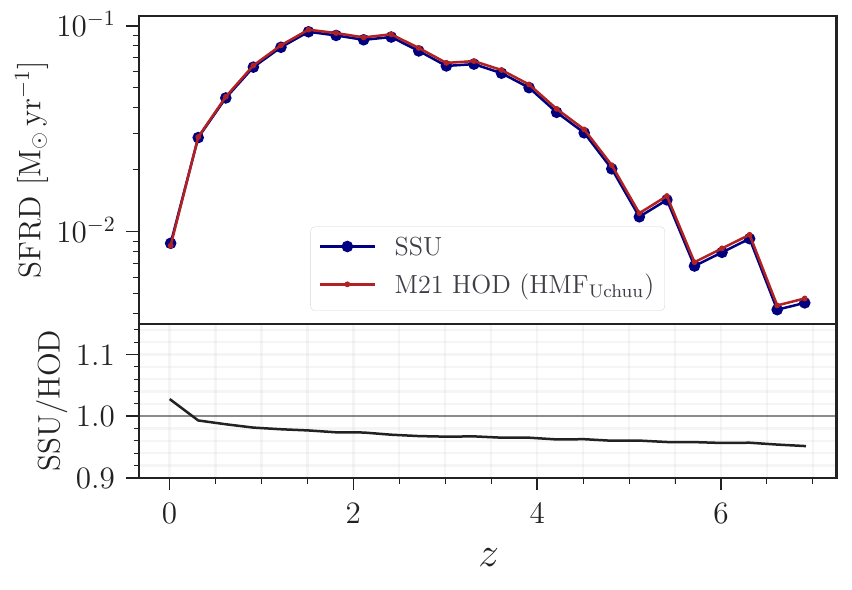}
    \caption{Comparison of the SFRD from the M21 HOD model (red line) and the SSU (blue line) as a function of redshift. The bottom panel displays the ratio of the SSU SFRD to that of the M21 HOD model.}
    \label{fig:1to1_SFRD}
\end{figure}

As shown in Eq.\,\ref{eq:diff_emissivity_centrals}, the SFR combines with the selected effective SED at a given frequency to construct the differential emissivity. We, thus, compare the SFRD of the M21 HOD model and the SSU as a function of redshift to determine whether they fully coincide, given that the same parameterization and assumptions for the SFR (Eq.\,\ref{eq:efficiency}) were applied. The result of the comparison is presented in Fig.\,\ref{fig:1to1_SFRD}.

For the SFRD of the HOD model, we first computed the SFR using Eq.\,\ref{eq:efficiency} at redshifts ranging from 0 to 7. Next, using the same HMF as the SSU, we computed the SFRD according to Eq.\,\ref{eq:HOD_SFRD_computation}
\begin{equation}
    {\rm SFRD}_{\rm HOD}(z) = \int \left.\frac{d^2N}{d{\rm log}M_h dV}(M_h, z)\right|_{\rm SSU} \times \, {\rm SFR}(M_h,z) \, d{\rm log}M_h.
\label{eq:HOD_SFRD_computation}
\end{equation} 
For the SFRD of the SSU, we binned the redshift range with centers corresponding to the same values used in the analytical calculation, assigning a bin width of $dz=0.5$ for each central redshift. We summed the SFR within each bin and divided by the corresponding comoving volume.

The two lines exhibit strong agreement across the entire redshift range. As indicated by the ratio of the lines, there is only a minor offset of less than 5\%, which is most noticeable at higher redshifts. This offset likely originates from numerical factors, such as the redshift binning required to compute the SFRD in the simulation.

\subsection{SSU vs HOD model: Differential emissivity $\left( \frac{dj}{dlogM_h} \right)$}
\label{subsec:dh_dlogMh_investigation}

The next quantity used to construct the power spectrum is the differential emissivity. For the halo model, we compute it using Eq.\,\ref{eq:diff_emissivity_centrals} over the redshift range from 0 to 7. In the simulation, the flux of each galaxy is calculated as $\chi^2 (1+z) \times \frac{SFR}{K} \times S_{\nu}^{\rm eff}$, where the redshift of each galaxy determines the comoving distance $\chi(z)$, the SFR via Eq.\,\ref{eq:efficiency}, and the effective SED at the corresponding frequency band. For a given redshift bin, we then bin galaxies by halo mass and sum the fluxes of all galaxies within the corresponding redshift and halo mass bins, dividing by the comoving volume of the redshift bin and the size of the halo mass bin.

\begin{figure}[h!]
    \centering
    \includegraphics[width=\linewidth]{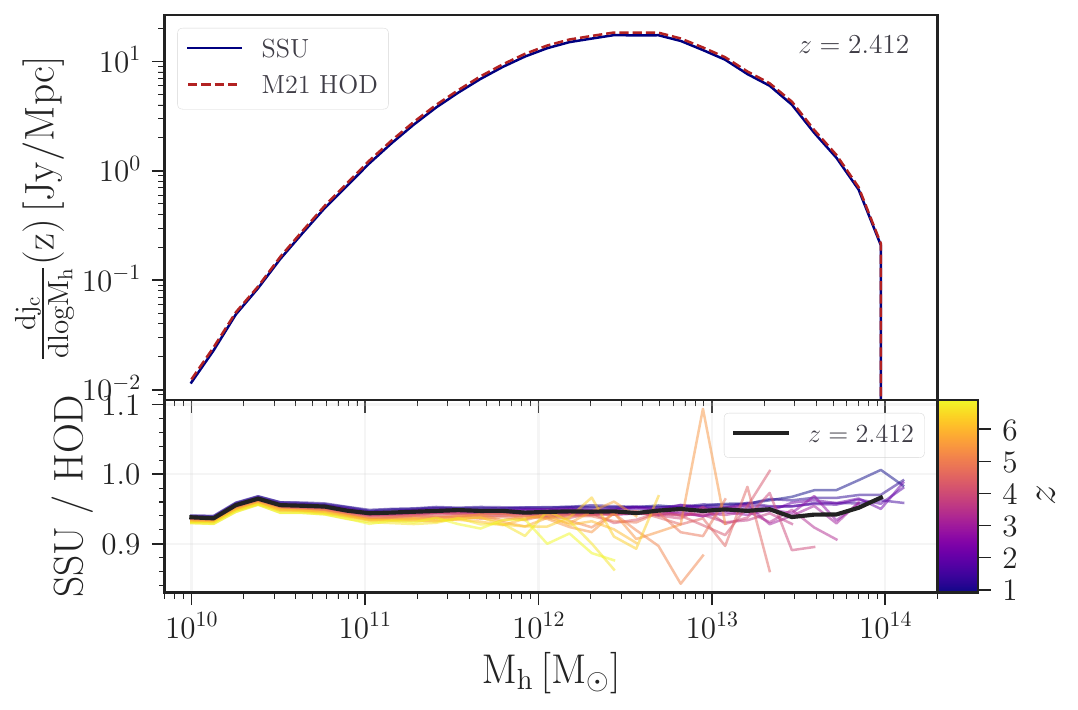}
    \caption{Comparison of the differential emissivity predicted by the M21 HOD model (red line) and the SSU model (blue line) at $z=2.412$. The bottom panel illustrates the ratio of the SSU model to the M21 HOD model at $z=2.412$ (black line), alongside the ratios for other redshifts (represented by lines in various colors).}
    \label{fig:1to1_dj_dlogMh}
\end{figure}

In the top panel of Fig.\,\ref{fig:1to1_dj_dlogMh}, the results for $z=2.412$ demonstrate very good agreement between the two lines. However, the ratio of the SSU model to the halo model, shown in the bottom panel of Fig.\,\ref{fig:1to1_dj_dlogMh}, reveals a slight offset (less than 5\%) across all halo masses. This offset is observed across all redshifts, with a slight increase at higher redshifts.

This offset arises from the need to discretize redshift when computing the analytical formulas. The effective SED ($S_{\nu}^{\rm eff}(z)$) as a function of redshift is continuous. In the simulation, each galaxy is assigned a $S_{\nu}^{\rm eff}(z)$ value based on its redshift, covering a very dense range of values along this curve. However, the analytical differential emissivity is computed at specific, equally spaced redshift intervals, corresponding to the given redshift grid. This means that all galaxies within a given redshift bin are assigned the same effective SED, based on the central value of the redshift bin. In contrast, galaxies in the simulation with redshifts close to each other but within the same bin are assigned slightly different $S_{\nu}^{\rm eff}(z)$ values based on their exact redshifts. These small discrepancies between the simulation and the analytical model accumulate within each redshift bin, leading to the observed systematic mismatch. Although this discrepancy is unavoidable, quantifying it helps us understand and predict how it will manifest in the final quantity of interest, the power spectrum.

\subsection{SSU Simulation vs halo model: Power spectrum}
\label{subsec:simu_vs_hod_pk}

The emission-related components, which are part of the HOD formalism, show strong agreement between the M21 HOD and the SSU, and even the small offset between the two has been quantified. We now turn to a direct comparison of the two-halo term of the power spectrum, which is the primary quantity we aim to model using the halo model formalism.

\begin{figure}[h!]
    \centering
    \includegraphics[width=\linewidth]{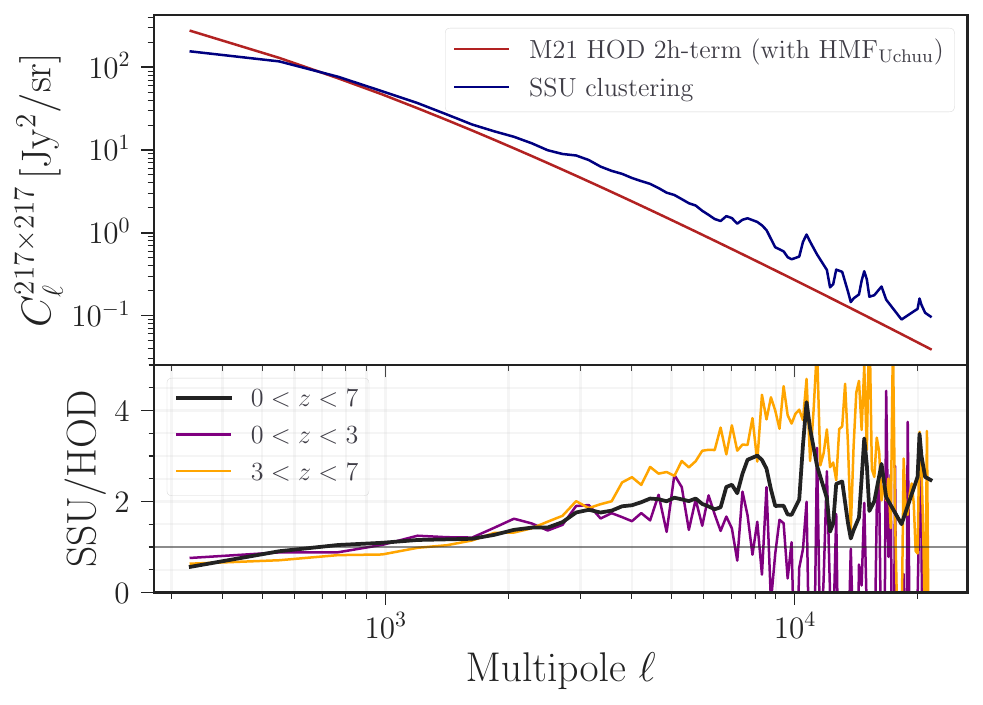}
    \caption{Comparison of the two-halo term from the M21 HOD model (red line) and the SSU (blue line), excluding subhalos. The bottom panel displays the ratio of the SSU to that of the M21 HOD model.}
    \label{fig:1to1_power_spectrum}
\end{figure}

The analytical computation of the two-halo term was performed using Eq.\,\ref{eq:two-halo_term}. For the simulation, we generated maps at all Planck frequencies from the SSU catalogs with subhalos excluded, and computed the power spectra from these subhalo-free maps. To obtain the clustering component, we subtracted the shot noise. Fig.\,\ref{fig:1to1_power_spectrum} shows the direct comparison of the analytical two-halo term and the clustering component of the simulation. There is a clear offset starting from the intermediate scales ($\ell > 2\times 10^3$) and increasingly extending down to smaller scales ($\ell > 10^4$). The small mismatch from the emission-related components cannot account for the amplitude of the mismatch observed in the power spectrum, nor can it explain its strong dependence on scale.

To better understand the source of this offset, we repeated the comparison between the two-halo term of the halo model and the simulation, focusing on two redshift bins: $0<z<3$ and $3<z<7$. These bins were chosen because sources at $z<3$ are expected to dominate the CIB \citep[e.g.,][]{bethermin2013_cib_z, schmidt2015}, while those at $z>3$ contribute significantly less. This division allows us to examine potential differences in the offset across these redshift ranges. The resulting ratios are presented in the bottom plot of Fig.\,\ref{fig:1to1_power_spectrum}. Although the power spectra become noisy at smaller scales, it is evident that the offset between the SSU and the M21 HOD model depends on scale. Interestingly, this scale-dependent offset appears more pronounced at $z>3$.  The fact that the discrepancy grows with increasing $\ell$ may help explain, at least in part, the mismatch between the intrinsic and recovered shot noise observed in the fitting process.

The next logical step is thus to investigate the dependencies of the remaining cosmological components in the power spectrum parameterization (Eq.\,\ref{eq:two-halo_term}), such as the bias model and the matter power spectrum. This investigation is particularly relevant given that the M21 HOD model employs a linear model for both the bias and the matter power spectrum.

\subsection{Discussion}
\label{subsec:other_sources_of_discrepancy} 

Our analysis shows that the emission-related components, which are governed by the HOD parameterization (e.g., SFR, emissivity) are in excellent agreement between the simulation and the analytical model. The differences are consistently below 5\% and well within expected numerical uncertainties. In contrast, the two-halo term shows a much larger discrepancy, exceeding 20\%. This points not to issues in the emission modeling but rather to limitations in the cosmological components of the broader halo model framework, such as the treatment of halo bias or matter clustering.

Specifically, when decomposing the two-halo term (Eq.~\ref{eq:two-halo_term}), the most likely contributors to the mismatch are the halo bias and matter power spectrum ingredients. Their product, $b^2 P_{\rm matter}$, sets both the amplitude and the scale dependence of clustering. In the M21 framework, both factors are taken from linear theory (bias from \citealt{tinker2010} and $P_{\rm matter}$ from \texttt{CAMB}), whereas SIDES–Uchuu includes the full nonlinear evolution of structure, naturally producing scale-dependent halo bias and a nonlinear matter power spectrum. This difference plausibly explains the redshift- and scale-dependent residuals we observe in the two-halo component.

Given these considerations, the discrepancy is best interpreted as arising from limitations in the overall halo modeling framework, rather than the HOD prescription alone. In particular, the use of linear approximations for halo bias and $P_{\rm matter}$ appears insufficient to match the fully evolved clustering signal of the simulation. This interpretation is consistent with other studies highlighting the breakdown of linear theory at intermediate scales and higher redshifts \citep{penin2018, acuto2021, mahony2022, dizgah2022, dvornik2023, jun2025}.

Future improvements to CIB modeling will likely require extending the halo model to include scale-dependent or nonlinear bias prescriptions (e.g., \citealt{fedeli2014, mead2021a, mead2021, dizgah2022}) and replacing the linear matter power spectrum with more accurate nonlinear models. These modifications would enhance the robustness of parameter inference, particularly as the precision of data increases.

\section{The effect of scatter on the SFR and SED}
\label{sect:scatter_effect}

The motivation to compare the M21 HOD model with a simplified version of the SIDES-Uchuu simulation came about when the model failed to accurately recover the exact parameters of the simulation. Therefore, the subsequent analysis presented in the previous sections, focused on the recipes governing the distribution and emission of sources (e.g., HMF, SFR). To isolate the effects of these recipes, we initially omitted any scatter in the relations used to generate the SFR and SED for each galaxy. In this section, we reintroduce this scatter to assess the extent of any disagreement that may arise from the SFR scatter, the SED scatter, or a combination of both.

\subsection{Estimating the impact of the SFR and SED scatter}
\label{subsec:how_sfr_sed_scatter_is_added}

The original SIDES-Uchuu simulation considers the SFR and SED scatter by construction. However, the way that the original simulation is built, it is rather complicated to switch on and off the scatter. In order to get around this, we use the SSU version of the simulation, where activating or deactivating the scatter in SFR or SED is straightforward. In this way we can quantify the impact of scatter's presence (or absence) on the final power spectrum. 

The SFR recipe used in the SSU is described by Eq.\,\ref{eq:efficiency}. Following the same approach as in the original SIDES-Uchuu, where a scatter of 0.36\,dex is used in the M$_{\rm h}$-SFR recipe (0.2\,dex scatter in the M$_{\rm h}$-M$_*$ relation through the abundance matching and 0.3\,dex in the M$_*$-SFR relation), we also used the same lognormal scatter into the SFR value of each source. However, introducing lognormal scatter requires an additional correction due to a property of the lognormal distribution. Its positive skewness shifts the mean to higher values compared to the central value. To ensure the mean SFR remains consistent with the case without scatter, we applied a correction to shift the mean SFR with scatter back to match the original value.

The M21 HOD model uses effective SEDs that depend only on the redshift of the sources, as detailed in Sect.\,\ref{subsect:simplified_SIDES}. Introducing scatter to the SEDs, however, allows each galaxy's SED template to be selected based on its mean radiation field value, $\langle U \rangle$, which in turn is determined by the galaxy's redshift with an added scatter of 0.2\,dex. As a result, galaxies at the same redshift can be assigned different SED templates. In contrast, when using effective SEDs, all galaxies at a given redshift share the same SED template. Furthermore, only main-sequence SED templates are employed, excluding starburst templates, as opposed to the original SIDES model. By enabling or disabling SED scatter, we aim to evaluate the impact of this simplification and assess the validity of the HOD model's approach. \\

We consider the four following cases:
\begin{itemize}
\item {\tt No Scatter}: Both SFR and SED scatter are disabled,
\item {\tt SFR Scatter}: SFR scatter enabled, SED scatter disabled,
\item {\tt SED Scatter}: SFR scatter disabled, SED scatter enabled,
\item {\tt SFR \& SED Scatter}: Both SFR and SED scatter are enabled.
\end{itemize}

The goal is to compare these four cases with each other, in order to quantify the extent to which each type of scatter influences the resultant power spectrum. We investigate this effect separately for the clustering and the shot-noise component of the power spectrum.

As previously, to calculate the shot noise, we generate maps by randomly rearranging the positions of the sources while keeping their luminosity unchanged, and then compute the power spectrum of these maps. To compute the clustering component, we subtract the shot noise spectrum from the total power spectrum.

\subsection{Results}

\begin{figure*}[h!]
    \centering
    \includegraphics[width=\linewidth]{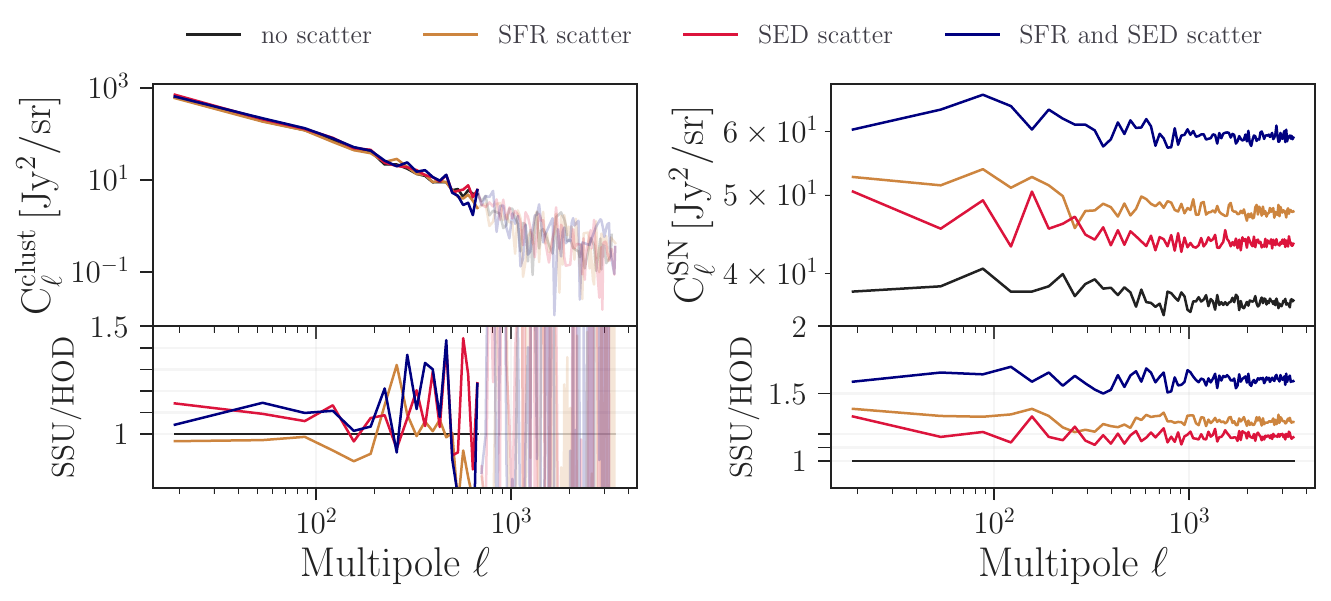}
    \caption{Power spectra computed using the simplified SIDES-Uchuu simulation for the 217$\times$217\,GHz \textit{Planck} frequency, featuring varying scatter scenarios, each represented by distinct colors, as detailed in the legend. The top panels display the raw power spectra, while the bottom panels depict the power spectrum ratios relative to the 'no scatter' case in linear scale. \textit{Left}: the clustering component of the power spectrum. \textit{Right}: the shot-noise component.}
    \label{fig:SED_SFR_scatter_effect_Pk}
\end{figure*}

Fig.\,\ref{fig:SED_SFR_scatter_effect_Pk} shows the power spectra at 217$\times$217\,GHz for both clustering and shot-noise components. The four different cases that result in slightly different power spectra, are indicated with different colors. Furthermore, the bottom panels feature a visual representation of the power spectrum ratios relative to the `no scatter' scenario, clearly demonstrating and quantifying the difference among the various cases. 

Introducing scatter significantly affects the shot-noise component, leading to noticeable differences in the power spectra compared to the `no scatter' case. Specifically, the SFR scatter results in a shift of $\sim 20\%$, while SED scatter alters the shot noise by $\sim 30\%$. Combining both SFR and SED scatter results in a power spectrum that deviates by over $50\%$ from the scenario without any scatter. This behavior is expected since the effect of the two different types of scatter adds up when used together. 

Things are different for the case of the clustering component, where it is obvious that the scatter has a smaller impact than in the shot noise. The addition of SFR scatter has an impact on the clustering below 5\%, while the SED scatter can introduce a shift of the order of $10\%$. 

\subsection{Discussion}

We anticipated that scatter in the emission relations would notably affect the shot noise, due to its high sensitivity to the flux of individual sources. The presence of scatter in either the SFR or the SED modifies the source luminosity, which directly affects their flux. Since the shot noise depends on an integral over the squared flux, even modest variations can substantially impact its amplitude. Typically, the shot noise of the CIB power spectrum provides insight into the abundance of faint, unresolved sources. However, our results indicate that the implementation of scatter in the relations of the M21 HOD model can lead to significantly different levels of shot noise. This variation could result in misinterpretations when analyzing observed shot noise amplitudes.

In contrast, the effect of scatter on the clustered (two-halo) component of the power spectrum is more modest in our study, with deviations remaining below 10\%. This result is consistent with the findings of \citet{jun2025b}, who studied the impact of secondary bias, specifically, correlations between star formation rate and halo structural properties, on the two-halo term of the LIM power spectrum. Using the IllustrisTNG simulation, they found that such correlations can enhance the two-halo term by up to 5\%. While their focus is on secondary environmental dependencies rather than stochastic scatter in emission relations, both mechanisms introduce similar scale-dependent modulations to the large-scale power. This strengthens the interpretation that SFR-related stochasticity and environmental effects may modestly bias clustering-based inferences, but are not dominant.

In comparison, \citet{murmu2023} found that scatter in the luminosity–halo mass relation can impact the [CII] power spectrum by up to 50\%. This is more aligned with our findings for the shot-noise component. However, a direct comparison is challenging: their study examined a single spectral line, whereas we analyze broadband continuum emission integrated over many sources and many redshifts. Moreover, \cite{murmu2023} focused on the total power spectrum, without disentangling the one-halo, two-halo, and shot-noise components, making it difficult to isolate which part is most affected. It is likely that the majority of the offset in their study stems from the shot-noise component, which aligns with our findings and is also consistent with \cite{gkogkou2023}, where it was shown that shot noise exhibits much greater variation for [CII] than for the CIB.

Overall, incorporating scatter in the emission relations of the HOD model would enhance its physical realism and versatility. However, scatter has only a limited impact on the clustered component and can reasonably be neglected when the focus is on large-scale structure modeling. On the other hand, it plays a critical role in the amplitude of the shot noise. Fully modeling scatter in the emission relations would significantly increase the complexity of the halo model framework, as simple mean scaling relations would no longer be sufficient. However, treating the shot noise as a free parameter offers a practical and effective alternative. It allows the model to absorb uncertainties introduced by scatter or sample variance \citep{gkogkou2023} without overcomplicating the parameter space.

\section{Summary and conclusions}
\label{sect:conclusions}

Halo models are a key tool for interpreting CIB anisotropies. They aim to extract astrophysical and cosmological parameters by linking galaxy emission to large-scale dark matter structure. In this work, we evaluated the four-parameter HOD model developed by \citet{maniyar2021}, which is embedded within the broader halo model framework and tailored for CIB data. Our goal was to test its ability to recover physical quantities by fitting it to simulated data with known ground truth.

We first fitted the M21 HOD model to mock data generated from the SIDES-Uchuu simulation, which includes a realistic treatment of galaxy formation and clustering. While the model reproduced the observed power spectra and SFRD reasonably well, it systematically failed to recover key intrinsic parameters, notably overestimating the halo mass corresponding to peak star formation efficiency. 

To isolate the cause, we constructed a simplified version of the simulation (SSU), adopting the same prescriptions as the M21 HOD model for SFR, SEDs, and halo occupation. Despite this alignment, best-fit parameters still deviated from the known inputs, suggesting that the problem lies beyond the choice of parameterization.

A detailed, one-to-one comparison between the HOD model and the SSU simulation revealed that the SFRD and differential emissivity agree to within 5\%, confirming that emission-related components are robustly modeled. However, we find a persistent, scale- and redshift-dependent offset in the two-halo component of the power spectrum, exceeding 20\%. This points not to shortcomings of the HOD itself, but to limitations of the broader halo-model ingredients, most notably the use of linear, scale-independent halo bias together with a linear matter power spectrum. These limitations are expected to be most prominent across the one- to two-halo transition, where nonlinear growth, scale dependent and higher order bias, and possible environmental dependencies become important. Addressing these effects, for example by incorporating scale dependent or nonlinear bias together with a nonlinear \(P(k)\), will be necessary for robust parameter recovery from high precision CIB data.

We also assessed the impact of including scatter in the SFR and SED assignments. This primarily affected the shot-noise component of the power spectrum (with shifts exceeding 50\%), while the clustered signal remained largely unchanged (below 10\% variation). This validates the treatment of shot noise as a free nuisance parameter but suggests that shot noise interpretations should be treated with caution when comparing to data.

In summary, although the halo model provides a good fit to the CIB power spectrum, it does not reliably recover the underlying physical parameters, even when matched to the exact simulation prescriptions. The dominant source of this discrepancy appears to lie in the broader halo model framework, particularly in its linear treatment of halo bias and matter clustering. Incorporating scale-dependent or nonlinear models for these cosmological components is likely necessary to improve the accuracy of halo model-based CIB analyses. This work underscores the value of simulation-based validation and has broader implications for both current CIB studies and future high redshift line-intensity mapping efforts.

\begin{acknowledgements}
  This project has received funding from the European Research Council (ERC) under the European Union’s Horizon 2020 research and innovation programme (grant agreement No 788212). 
  This work was supported by the TITAN ERA Chair project (contract no. 101086741) within the Horizon Europe Framework Program of the European Commission.
   We thank M. Jose Luis Bernal, Azadeh Moradinezhad Dizgah, and Mathilde Van Cuyck for the helpful discussions.
\end{acknowledgements}

\bibliographystyle{aa}

\appendix

\section{Covariance matrix estimation}
\label{ap:covariance_matrix}

To accurately estimate uncertainties and correlations between the power spectra at different frequencies and angular scales, we computed the full Gaussian covariance matrix using the NaMaster flat-sky formalism. This was done across the four \textit{Planck} frequencies (217, 353, 545, and 857 GHz) and eight multipole bins (see Fig.\,\ref{fig:fit2realSIDES_Pk}). The computation uses simulated maps covering the full SIDES-Uchuu area with a uniform mask, under the flat-sky approximation.

NaMaster internally accounts for mode coupling via the coupling matrix derived from the sky mask. However, due to the large and nearly uniform sky coverage of the SIDES-Uchuu maps, we find that off-diagonal terms from mode coupling are negligible. Therefore, while mode coupling is included in the computation, its impact is minimal in our case.

\begin{figure}[h!]
    \centering
    \includegraphics[width=\linewidth]{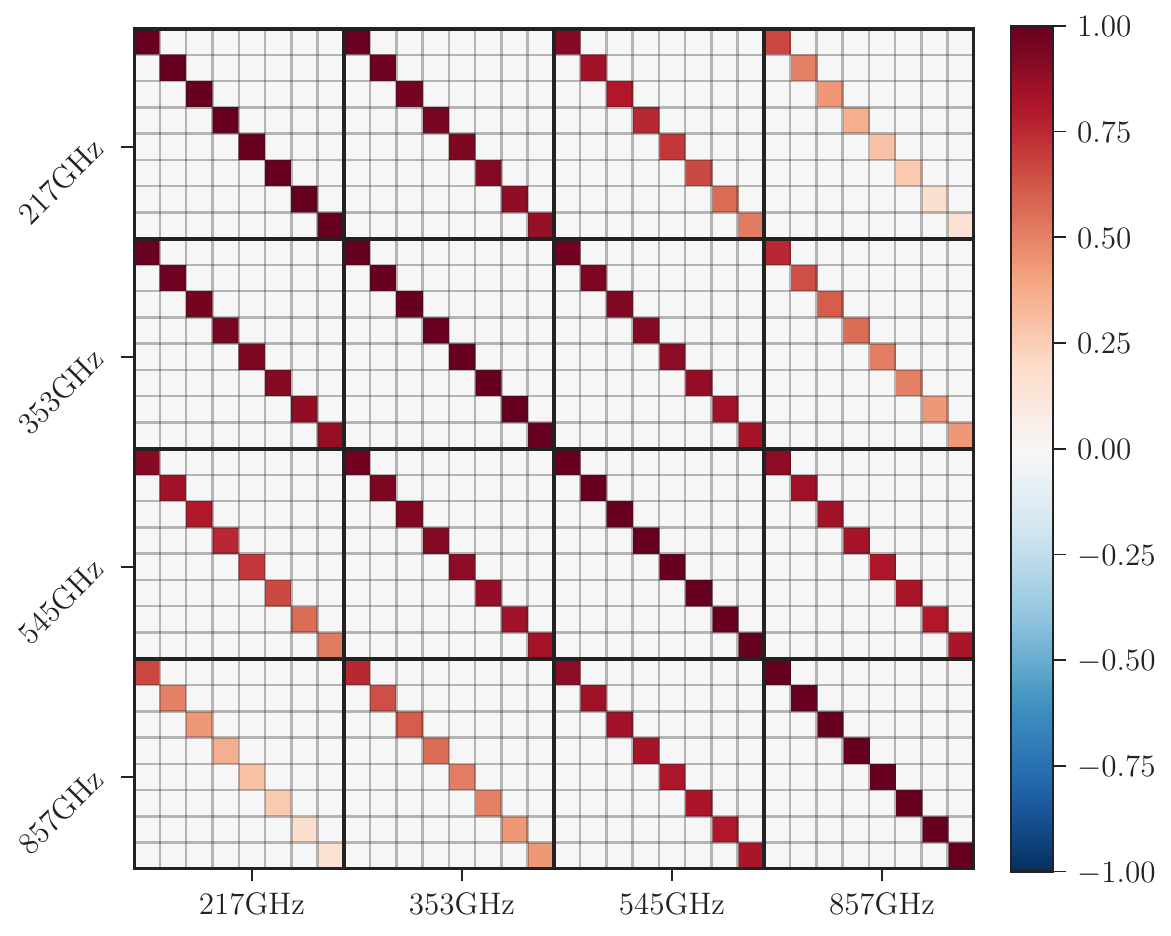}
    \caption{Correlation matrix of the SIDES-Uchuu power spectra for the four \textit{Planck} frequency bands at 8 $\ell$ bins.}
    \label{fig:covmat_realSIDES}
\end{figure}

\begin{figure*}[h!]
    \centering
    \begin{minipage}{0.45\textwidth}
        \centering
        \includegraphics[width=\textwidth]{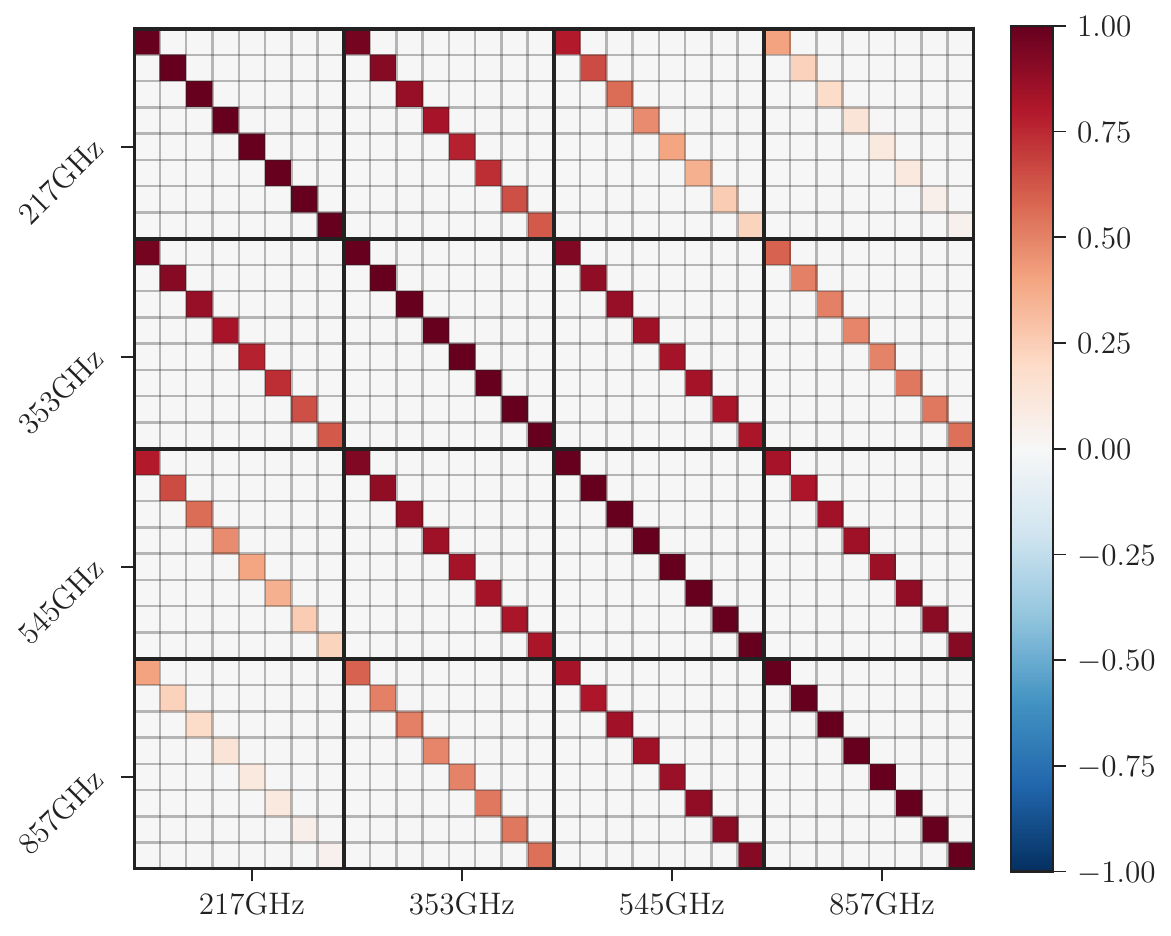}
    \end{minipage}
    \hfill
    \begin{minipage}{0.45\textwidth}
        \centering
        \includegraphics[width=\textwidth]{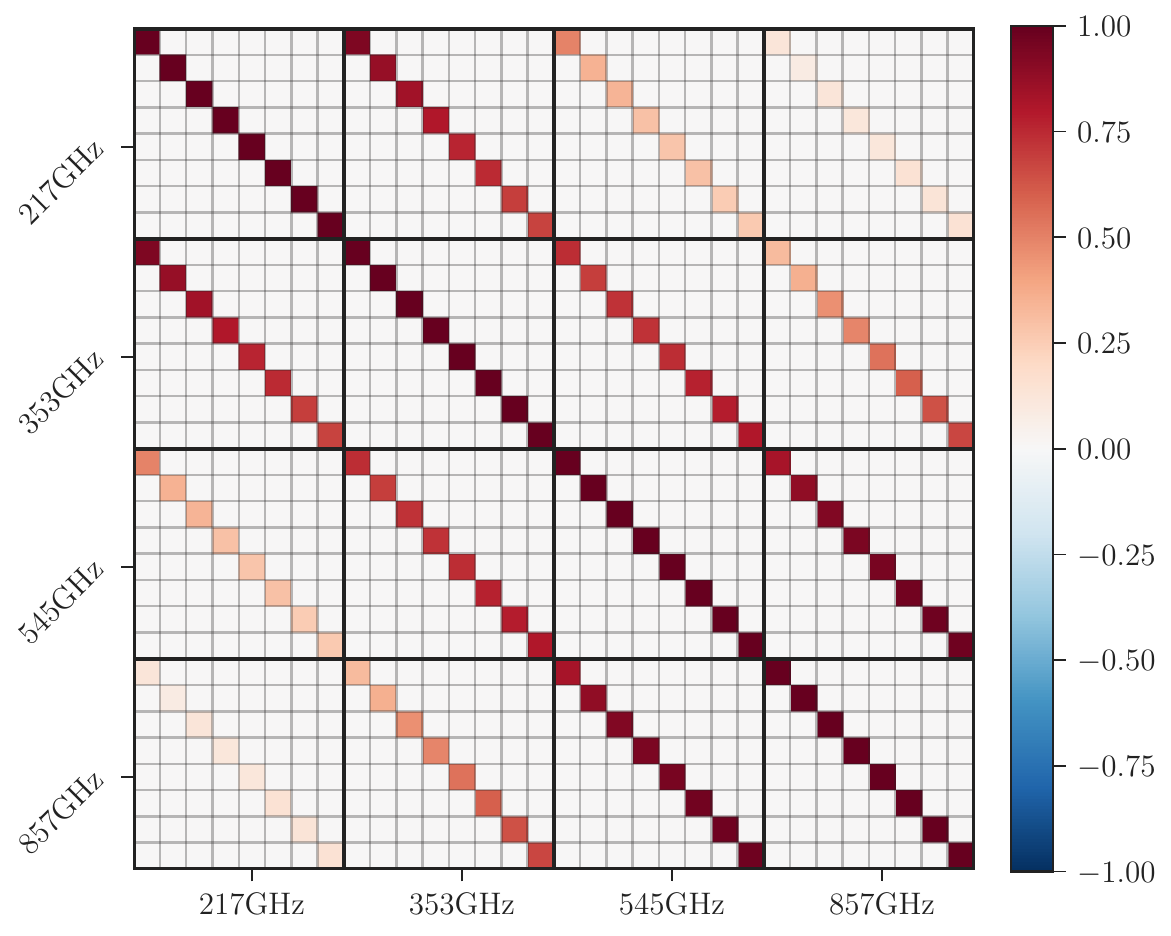}
    \end{minipage}

    \vspace{0.2cm}

    \begin{minipage}{0.45\textwidth}
        \centering
        \includegraphics[width=\textwidth]{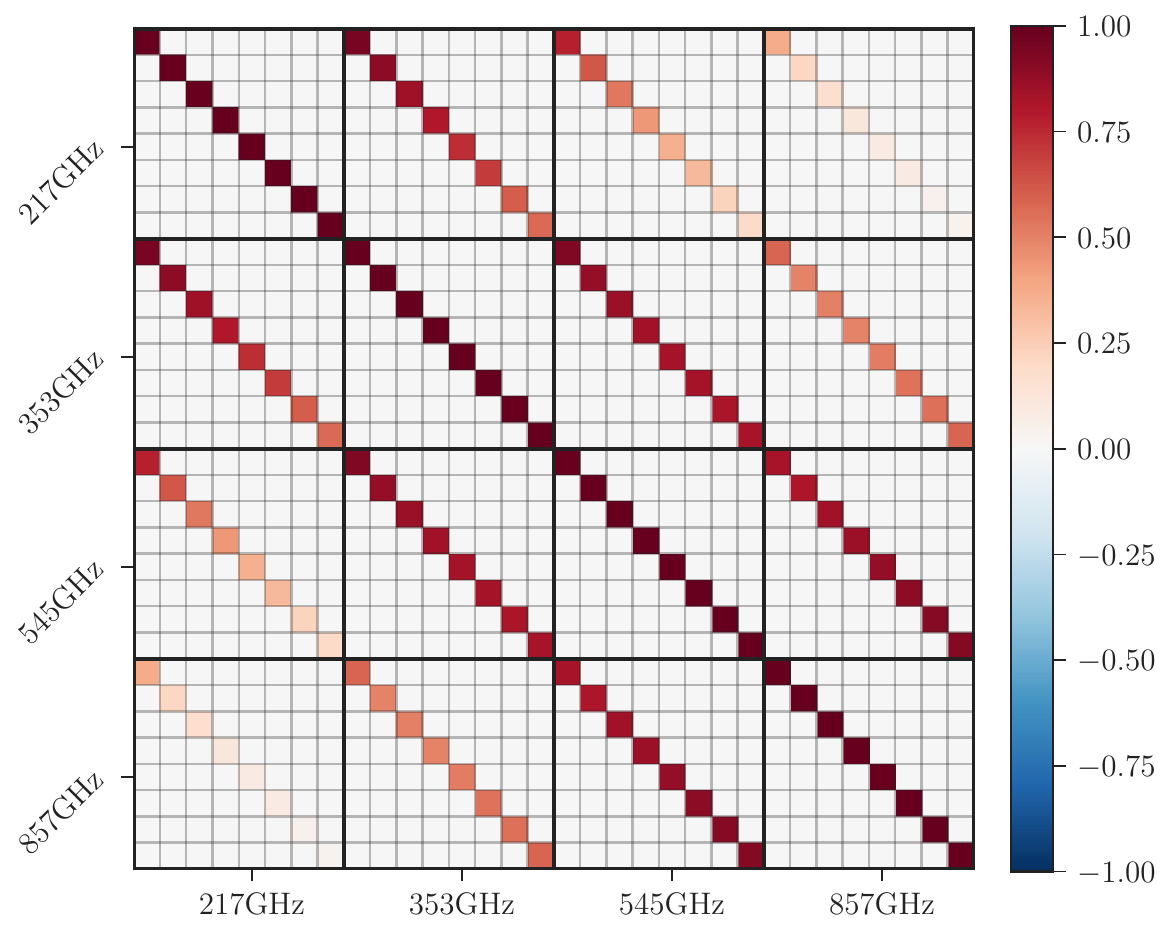}
    \end{minipage}
    \hfill
    \begin{minipage}{0.45\textwidth}
        \centering
        \includegraphics[width=\textwidth]{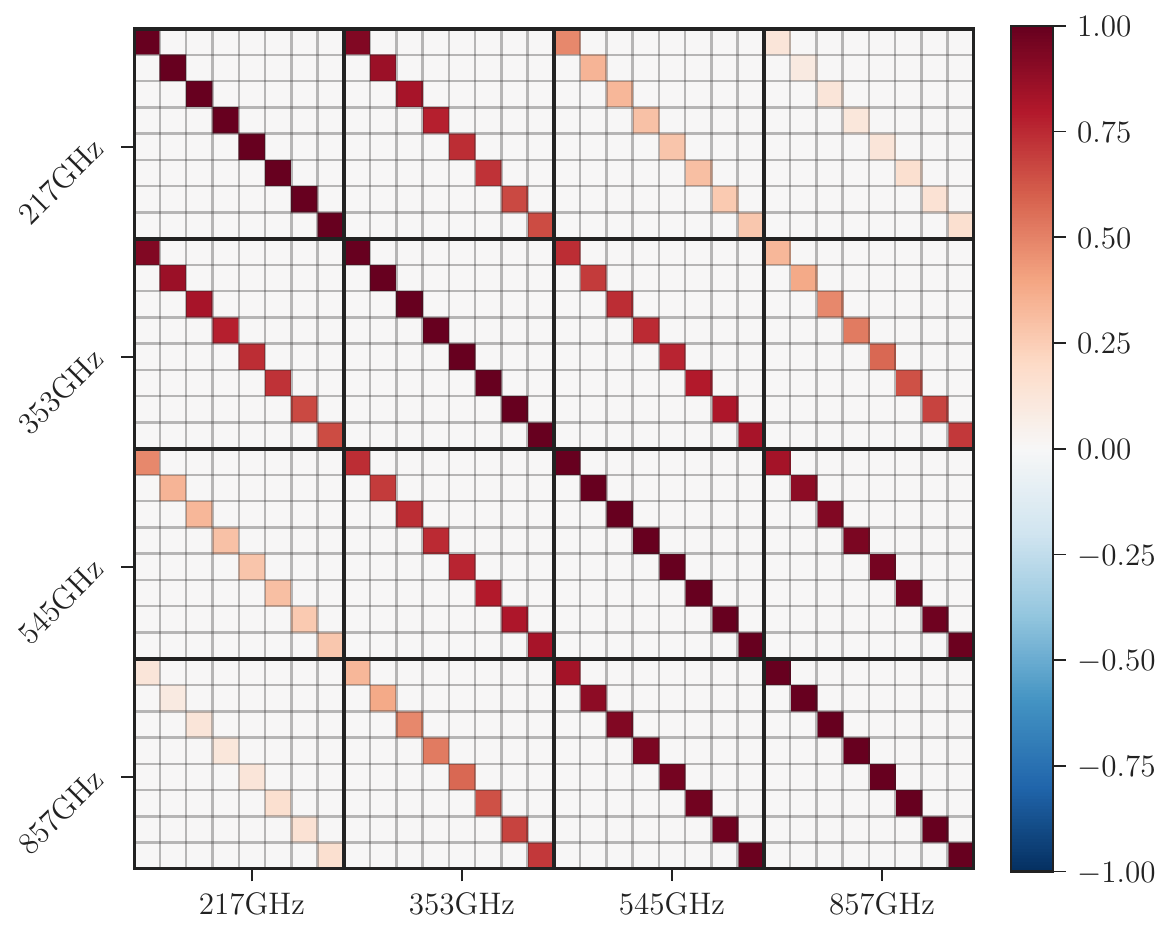}
    \end{minipage}
    \caption{Correlation matrices of the four \textit{Planck} frequency bands for the four different SSU configurations. Top row, left to right: 'with subs' and $\tau \ne 0$, 'with subs' and $\tau = 0$. Bottom row, left to right: 'no subs' and $\tau \ne 0$, 'no subs' and $\tau = 0$.}
    \label{fig:covmats_SSU}
\end{figure*}

The resulting $32 \times 32$ matrix captures the joint covariance structure of all power spectra. The square roots of its diagonal elements are used as error bars in the analysis. This procedure is repeated for each simulation configuration considered in this work. The corresponding correlation matrices for the SIDES-Uchuu and SSU cases are shown in Fig.\,\ref{fig:covmat_realSIDES} and Fig.\,\ref{fig:covmats_SSU}, respectively. The correlation matrix $R_{i,j}$ is computed from the covariance matrix $C_{i,j}$ as:
\begin{equation}
R_{i,j} = \frac{C_{i,j}}{\sqrt{C_{i,i} \times C_{j,j}}}
\label{eq:correlation_matrix_from_covmat}
\end{equation}

\section{Fitting the M21 HOD model to SSU data with $\tau = 0$}
\label{ap:tau_off_results}

In this section we show the results of fitting the M21 HOD model to the SSU data, with the $\tau$ parameter (which regulates the evolution of the efficiency curve width with redshift) set to zero. We explored this scenario to evaluate whether a simpler parameterization of the efficiency in converting baryonic accretion into star formation could improve the M21 HOD model's ability to describe the simulated SSU data.

\begin{figure*}[h!]
    \centering
    \includegraphics[width=0.8\linewidth]{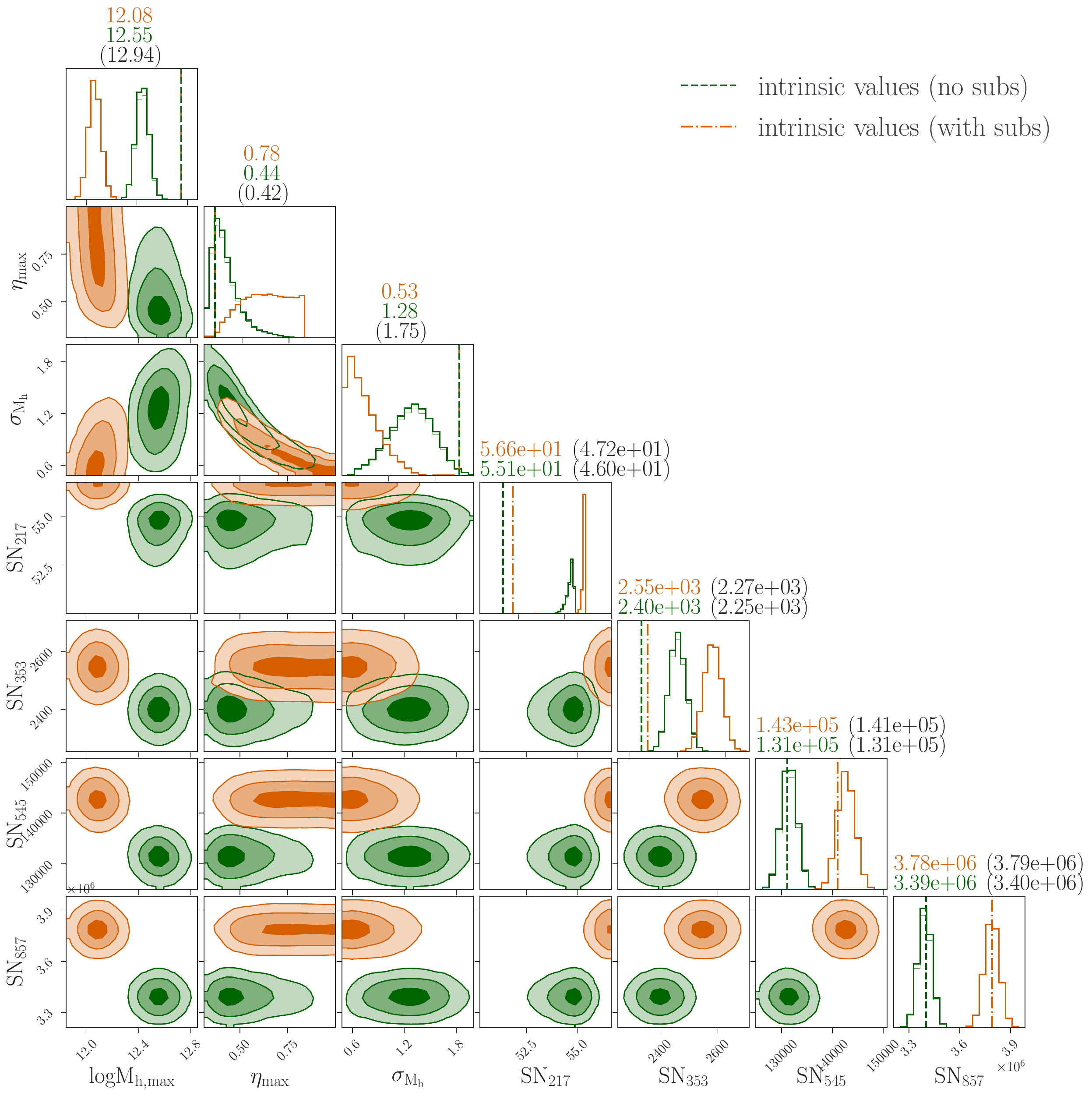}
    \caption{Resulting corner plots from fitting the M21 model to mock data generated with SSU for the case of $\tau$=0 no evolution of the width of the efficiency curve with respect to redshift). The orange contours and histograms correspond to "with subs", while the green color represents the "no subs" case.}
    \label{fig:corner_plot_simplified_with_vs_no_subs_tauOFF}
\end{figure*}

\begin{figure*}[h!]
    \centering
    \includegraphics[width=0.6\linewidth]{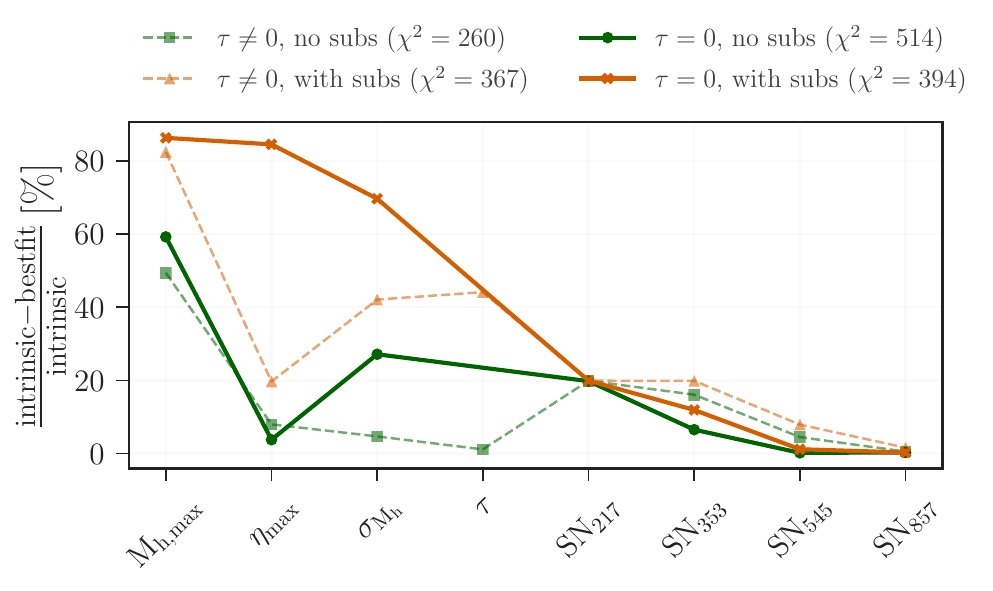}
    \caption{Percentage difference between the intrinsic and the best-fit values of the model parameters for the case of $\tau = 0$ while keeping the $\tau \neq 0$ case from Fig.\,\ref{fig:perecentage_difference_tauON} for reference.}
    \label{fig:perc_difference_with_vs_no_subs_tauOFF}
\end{figure*}

The fitting procedure follows the same methodology described in Sect.\,\ref{sect:hod_vs_simplifiedSIDES}. The resulting corner plot is displayed in Fig.\,\ref{fig:corner_plot_simplified_with_vs_no_subs_tauOFF} and the percentage differences between the intrinsic simulation values and those derived from the HOD model are shown in Fig.\,\ref{fig:perc_difference_with_vs_no_subs_tauOFF}. All parameters are recovered with comparable accuracy to the $\tau \neq 0$ case, showing no significant improvement. This result is unexpected, as removing $\tau$ was intended to break the degeneracy with $\sigma_{\rm M_h}$ and thereby improve parameter recovery.

The aim of this exercise was not to advocate for a simpler parameterization, but rather to evaluate whether this modification could eliminate the observed offsets and clarify their origin. Since the offsets persist, the results suggest that they likely stem from a more fundamental limitation within the components of the M21 HOD model.

\section{Complementary plots}
\label{ap:complementary_plots}

\begin{figure*}[h!]
    \centering
    \includegraphics[width=0.8\linewidth]{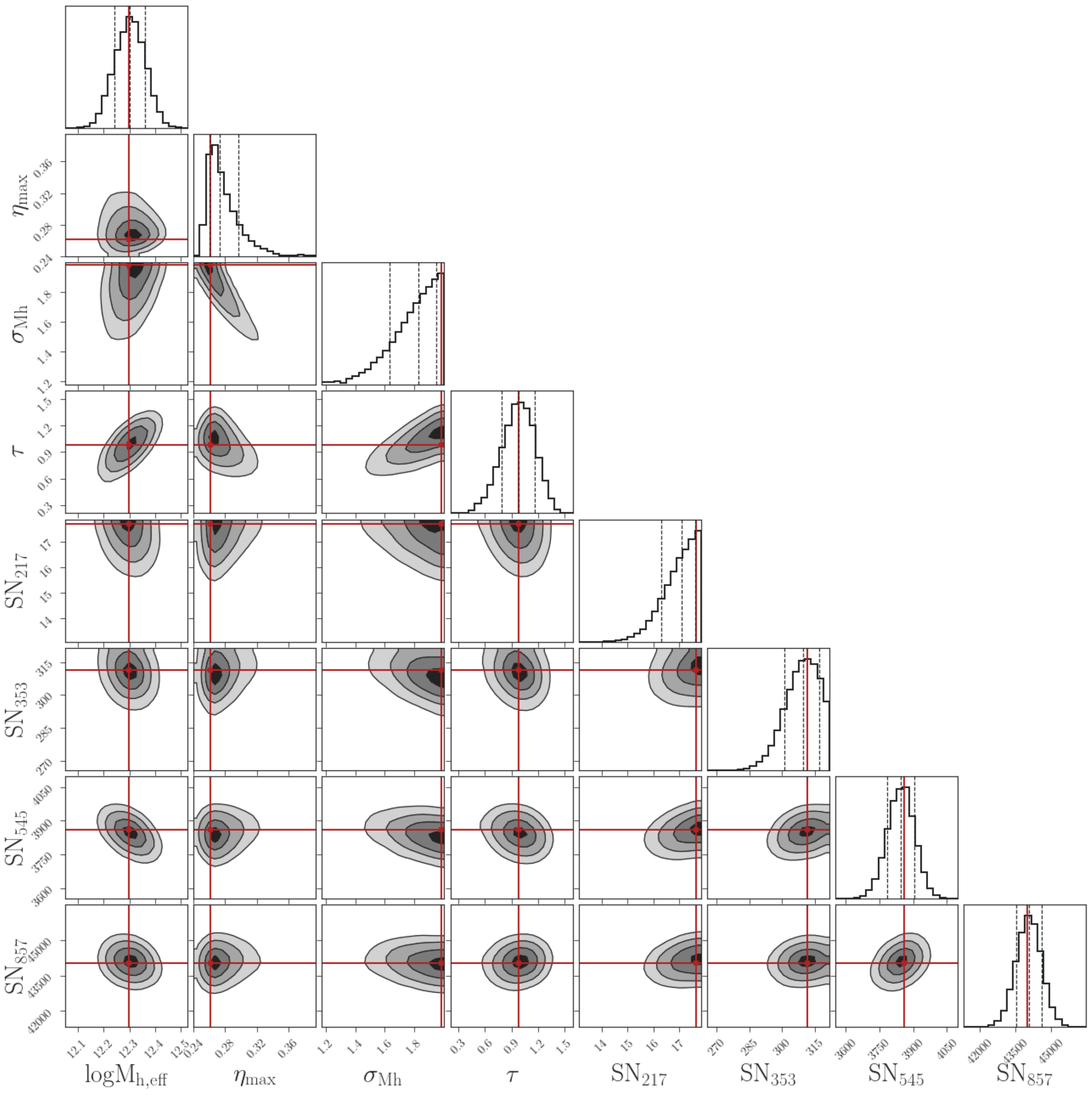}
    \caption{Corner plots depicting the outcomes of fitting the M21 HOD model to mock SIDES data. The best-fit values of the model parameters are highlighted in red.}
    \label{fig:fit2realSIDES_cornerplot}
\end{figure*}

Fig.\,\ref{fig:fit2realSIDES_cornerplot} presents the corner plots obtained from fitting the M21 HOD model to simulated data from the original SIDES-Uchuu simulation. Details of the fitting process are provided in Sect.\,\ref{sect:fit2realSIDES}.

\begin{figure*}[h!]
    \centering
    \includegraphics[width=0.6\linewidth]{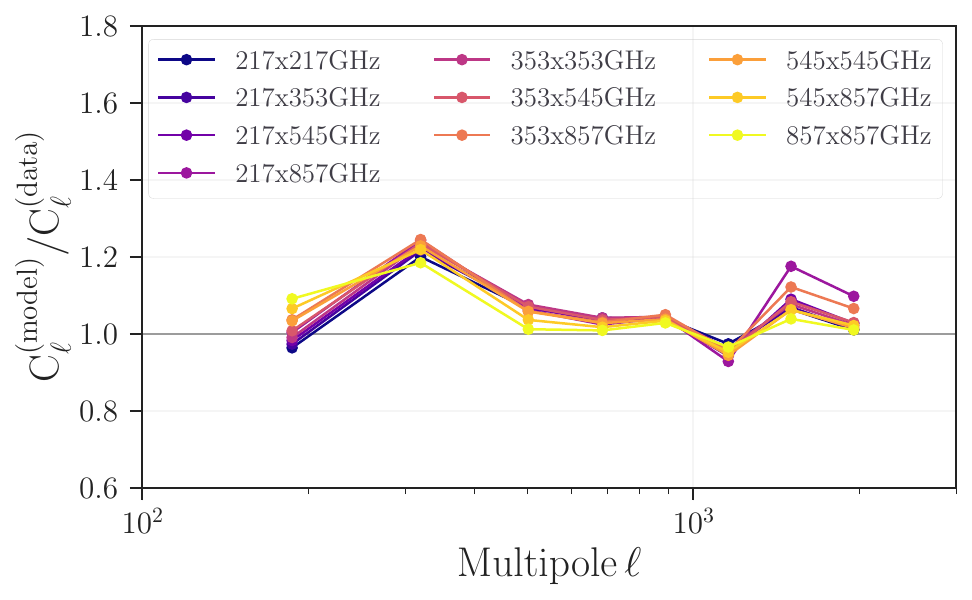}
    \caption{Ratio of the CIB halo model over the mock data obtained with the original SIDES (Fig.\,\ref{fig:fit2realSIDES_Pk}). The different colors correspond to different frequency pairs.}
    \label{fig:fit2realSIDES_power_spectra_ratios}
\end{figure*}

To complement our results, we include additional plots for all other frequencies analyzed in our tests. Fig.\,\ref{fig:fit2realSIDES_power_spectra_ratios} shows the ratio of the original SIDES-Uchuu simulated data to the power spectra derived from the HOD model's best-fit parameters for all combinations of the \textit{Planck} frequency bands. The ratios exhibit the same behavior across all frequencies as presented in the main text for the $217 \times 217$ GHz case, with only minor variations in amplitude.

\begin{figure*}[h!]
    \centering
    \includegraphics[width=0.8\linewidth]{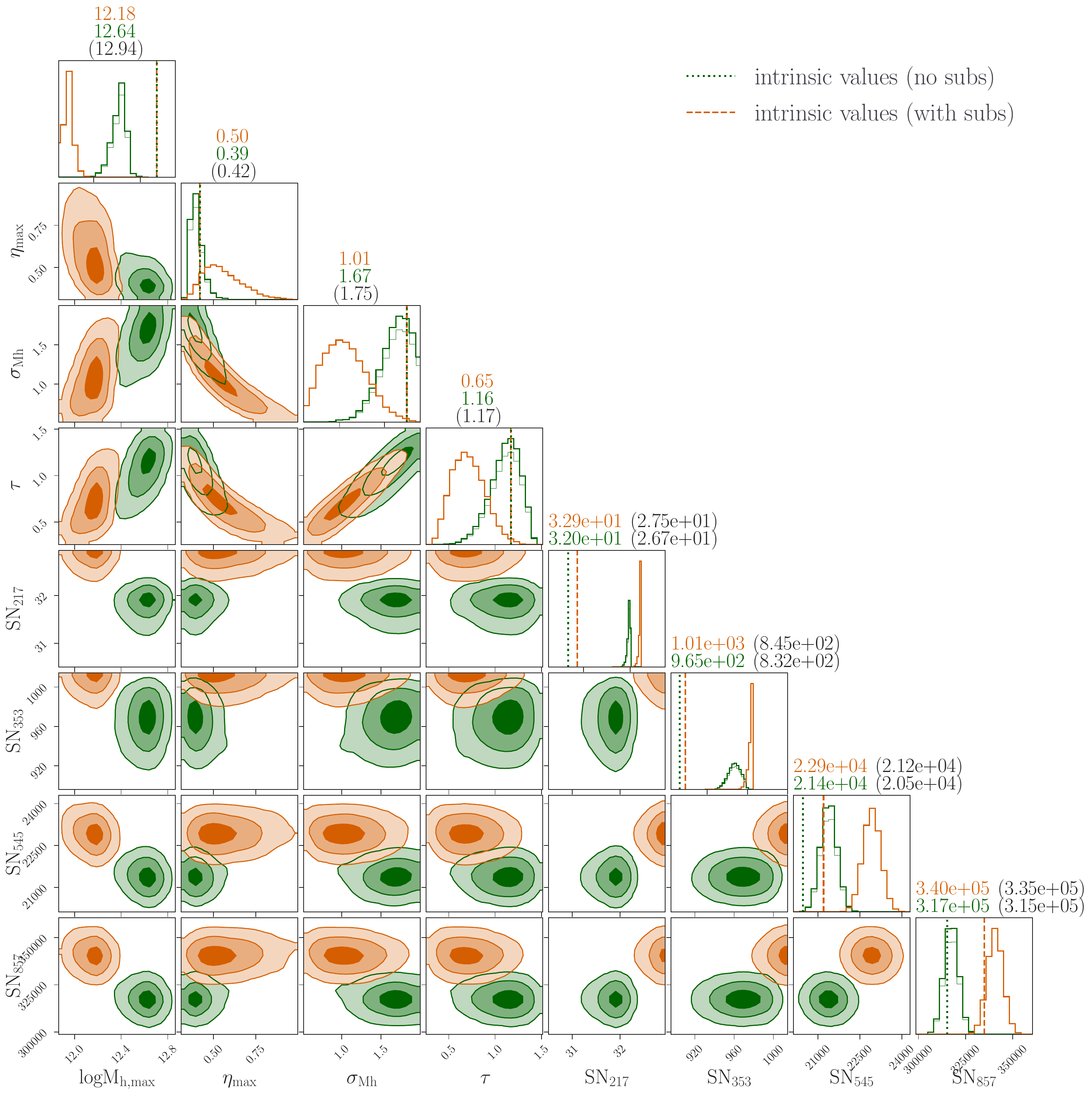}
    \caption{Resulting corner plots from fitting the M21 model to mock data generated with SSU (see Sect.\,\ref{subsect:simplified_SIDES} for a detailed description). The orange contours and histograms correspond to the scenario where the subhalos are considered ("with subs"), while the green color represents the case when subhalos are excluded ("no subs"). The intrinsic parameter values for the "with subs" and "no subs" cases are illustrated by the orange dotted dashed line and the green dashed line, respectively. Above each histogram, the best-fit value is displayed in the corresponding color, while the intrinsic value is shown in parentheses and in black. Note that the intrinsic values of the shot noise in the "no subs" and "with subs" case are not the same.}
    \label{fig:corner_plot_simplified_with_vs_no_subs_tauON}
\end{figure*}

Fig.\,\ref{fig:corner_plot_simplified_with_vs_no_subs_tauON} presents the corner plots obtained from fitting the M21 HOD model to SSU simulated data in the case of "with subs" and $\tau \neq 0$. Details of the data generation and fitting process are provided in Sect.\,\ref{sect:hod_vs_simplifiedSIDES}.

\end{document}